\documentstyle [eqsecnum,preprint,aps,epsf,fixes]{revtex}

\def\eps{\epsilon}

\def\eff{{\rm eff}}

\def\meff{m_\eff}
\def\nc{n_{\rm c}}
\def\bare{{\rm bare}}
\def\betaone{\beta^{(1)}}
\def\betatwo{\beta^{(2)}}

\def\gammatwo{\gamma^{(2)}}
\def\MSbar{$\overline{\rm MS}$}
\def\tr{{\rm tr}}
\def\asym{{\rm asym}}
\def\heavy{{\rm heavy}}
\def\light{{\rm light}}
\def\half{{\textstyle {1\over2}}}
\def\On{O($n$)}
\def\figrule{\vrule width 0pt}

\def\nnorm{\kappa}
\def\norm{{\cal N}}
\def\normu{\norm\!\mu^\eps}

\def\ma{m_{\rm a}}
\def\mb{m_{\rm b}}

\def\vconst{C}
\def\vconsta{\vconst_{11}}
\def\vconstb{\vconst_{10}}
\def\vconstc{\vconst_{\eps10}}
\def\vconstcm{\vconst_{\eps m10}}
\def\vconstd{\vconst_{\eps\eps10}}

\begin {document}

\preprint {UW/PT-96-23}

\title {$\eps$ expansion analysis of very weak first-order transitions\\
      in the cubic anisotropy model: Part I}

\author {Peter Arnold and Laurence G.~Yaffe}

\address
    {%
    Department of Physics,
    University of Washington,
    Seattle, Washington 98195
    }%
\date {October 22, 1996}
\maketitle

\begin {abstract}%
{%
  The cubic anisotropy model
  provides a simple example of a system with an arbitrarily weak first-order
  phase transition.
  We present an analysis of this model using $\eps$-expansion techniques
  with results up to next-to-next-to-leading order in $\eps$.
  Specifically, we compute the
  relative discontinuity of various physical quantities
  across the transition in the limit that the transition becomes arbitrarily
  weakly first-order.
  This provides a useful test-bed for the application
  of the $\eps$ expansion in weakly first-order transitions.
\ifpreprintsty
\thispagestyle {empty}
\newpage
\thispagestyle {empty}
\vbox to \vsize
    {%
    \vfill \baselineskip .28cm \par \font\tinyrm=cmr7 \tinyrm \noindent
    \narrower
    This report was prepared as an account of work sponsored by the
    United States Government.
    Neither the United States nor the United States Department of Energy,
    nor any of their employees, nor any of their contractors,
    subcontractors, or their employees, makes any warranty,
    express or implied, or assumes any legal liability or
    responsibility for the product or process disclosed,
    or represents that its use would not infringe privately-owned rights.
    By acceptance of this article, the publisher and/or recipient
    acknowledges the U.S.~Government's right to retain a non-exclusive,
    royalty-free license in and to any copyright covering this paper.%
    }%
\fi
}%
\end {abstract}

\section {Introduction}

The cubic anisotropy model is a simple two-scalar model that, for a certain
range of parameters, has a phase transition
with similarities to the finite-temperature phase
transition of electroweak theory in the early universe.
More generally, it provides a simple prototype for systems that have
weak, fluctuation-induced, first-order phase transitions
\cite{rudnick,amit}.
In ref.~\cite{summary}, we and our collaborators discuss in detail
the similarities and dissimilarities with the electroweak transition
(noted earlier by Alford and March-Russel \cite {alford})
and described how the cubic
anisotropy model is a good testing ground for
analytic techniques that claim to distinguish between
second-order and weakly first-order phase transitions.
In particular, the model can be used to study $\eps$ expansion methods, which
we have previously applied to the electroweak case \cite{arnold&yaffe1}.
This paper presents the details of calculating
$\eps$ expansions for weakly first-order phase transitions in
the cubic anisotropy model.
A better overview of the motivation, a summary of our $\eps$ expansion
results, and a comparison against numerical Monte Carlo simulations
\cite {numerical}
may be
found in ref.~\cite{summary}.

Our goal will be to compute the relative discontinuity of various quantities
(the specific heat, susceptibility, and correlation length) across the
phase transition when one has an extremely weak first-order transition.
Specifically, we will compute ratios such as $\chi_+/\chi_-$ where
$\chi_\pm$ are the susceptibilities on either side of the transition.
This was originally done at leading order in $\eps$ by Rudnick some
twenty years ago \cite{rudnick}.%
\footnote{
   Our leading-order results for
   $\chi_+/\chi_-$ and $C_+/C_-$, however, differ by factors of 4
   from ref.~\cite{rudnick}.
}
We have extended the calculation to
next-to-leading order (NLO) for the correlation length and
next-to-next-to-leading order (NNLO) for the susceptibility ratio.
A next-to-leading order
calculation for the specific heat ratio is performed in
a companion paper \cite{specific-heat}.


\subsection{Three-dimensional reduction and the cubic anisotropy model}

Before introducing the cubic anisotropy model, we shall very briefly
review the connection between phase transitions in thermal quantum
field theory and those in classical statistical mechanics.  For definiteness,
consider the topical example of electroweak theory.
In studying the electroweak transition, one starts with a 3+1 dimensional
SU(2) gauge-Higgs theory at finite temperature.%
\footnote
    {%
    Fermions, and the U(1) and SU(3) gauge fields, do not have a major impact
    on the phase transition dynamics and are, for simplicity, neglected.
    }
Schematically, the
Euclidean action is of the form
\begin {equation}
   S = \int_0^\beta d\tau \int d^dx \>
   \left\{ {\displaystyle {1\over2}} \, |D\phi|^2
       + {\displaystyle {1\over4}} \, F^2
       - {\displaystyle {\mu^2\over2}} \, |\phi|^2
       + {\displaystyle {\lambda\over4!}} \, |\phi|^4
       + \cdots
   \right\} \,,
\end {equation}
(with gauge-fixing terms omitted).
$\beta$ is the inverse temperature, and
$d=3$ is the number of spatial dimensions.
If the correlation length at the transition is large compared to the
inverse temperature (which is generally the case), one may simplify the
study of equilibrium properties of the transition by integrating out
the dynamics of the Euclidean time direction.  This yields
an effective three-dimensional theory that describes the long distance
physics of the transition and which may be precisely matched, order by
order in coupling constants, to the original theory:%
\footnote{
   Another way of explaining the appearance of a three-dimensional theory
  is to note that, if effective particle masses are small compared to $T$
  at the transition, then for small momenta
  the Bose distribution function $1/(e^{\beta E}{-}1)$
  is large compared to one.
  But physics should be classical
  if the number of quanta in each state is large.  The long-distance physics
  of the transition can therefore be approximated by classical statistical
  mechanics in three spatial dimensions.  By the well-known equivalence of
  statistical mechanics and quantum mechanics, this is equivalent to a
  ``zero-temperature'' field theory in three Euclidean space-time dimensions,
  which is one way to view the effective theory (\ref{Seff EW}).
}
\begin {equation}
   S_\eff = \beta \int d^dx \>
       \left\{ {1\over2} \, |D\phi|^2
        + {1\over4} \, F^2
        + {t \over 2} \, |\phi|^2		
	+ {\lambda_{\rm eff} \over4!} \, |\phi|^4
	+ \cdots
	\right\} \,.
\label{Seff EW}
\end {equation}
For a review, see refs.~\cite{arnold&yaffe1,3d}.
There is no need to go into detail here, except to note that the
mass-squared $t$ of the Higgs in the effective theory has the form
\begin {equation}
   t = -\mu^2 + c \, g^2 T^2 + \cdots \,,
\end {equation}
for some constant $c$.
The fact that $t$ becomes positive as the temperature $T$ increases
drives the restoration of manifest SU(2) symmetry at high
temperature.
The action (\ref{Seff EW}) describes a classical statistical mechanics
problem in three spatial dimensions, where $S_\eff$ is to be interpreted
as $\beta H$.

In this work, we will not study the three-dimensional theory (\ref{Seff EW}),
but will instead examine a simpler three-dimensional theory consisting of two
scalar fields known as the cubic anisotropy model.
In a more general form, the cubic anisotropy model is an \On\ symmetric
scalar model of $n$ real scalar fields, to which is added
an interaction that breaks \On\ symmetry down
to hyper-cubic symmetry
\cite{rudnick,amit}.
The action is
\begin {equation}
   S = \int d^dx \> \left\{ {1\over2}\left|\partial\vec\phi\right|^2 
                 + {t\over2} \left|\vec\phi\right|^2
                 + \nnorm u \left|\vec\phi\right|^4
                 + \nnorm v \sum_i \phi_i^4
       \right\} \,.
\label{eq:action0}
\end {equation}
(Note that the overall $\beta$ in (\ref{Seff EW}) can be absorbed by a
rescaling of $\phi$.)
We will ultimately be interested in the simplest case, $n=2$.
The parameters
$u$ and $v$ are dimensionless coupling constants, and $\nnorm$ is a
dimensionful normalization which we will fix later.
The phase transition of interest occurs as the parameter $t$ is varied.
At tree level,%
\footnote{
  There is some ambiguity of language depending on whether one views
  the action (\ref{eq:action0}) as (a) describing classical statistical
  mechanics of a field theory in $d{=}3$ spatial dimensions,
  with $S$ equaling $\beta H$, or
  (b) as a quantum-field theory in $d{=}3$ Euclidean space-time
  dimensions.
  In the former case, a ``tree level'' result would normally be referred to as
  a ``mean field theory'' result; in the latter, it would be referred to
  as a ``classical'' result.  The first interpretation more accurately
  reflects the physics of the problem, but the latter is more familiar
  to particle theorists.  We shall bypass the terminology
  issue by referring simply to tree-level 
  vs.\ one-loop results, {\em etc}. 
}
the transition appears to be second-order,
with hyper-cubic symmetry spontaneously broken for $t<0$ and restored
for $t>0$.  As we shall review, however, the effect of higher-order
corrections on the nature of the transition cannot be ignored.


\subsection{The $\eps$ expansion}

In second-order phase transitions, the correlation length diverges
at the transition and physics near the transition is dominated by large
infrared fluctuations which cannot be treated perturbatively.
One way of summarizing this for a theory like the cubic anisotropy model
is to note that, since the couplings $\nnorm u$ and $\nnorm v$
of the theory have
non-trivial mass dimension $4{-}d = 1$, the dimensionless loop expansion
parameter $R$ must, by dimensional analysis, be of the form
\begin {equation}
   R ~~\sim~~ \hbox{($\nnorm u$ or $\nnorm v$)}~~\times~~
        \hbox{(some correlation length)}^{4-d} \,,
\end {equation}
which diverges with the correlation length.  In first-order phase
transitions, the correlation length is finite, but perturbation theory
can still fail if the correlation length is large enough that $R \gtrsim 1$.
We shall refer to such a situation as a weakly first-order transition.
In this paper, our goal will be to study the arbitrarily weak limit
$R \to \infty$.

The $\epsilon$ expansion is based on generalizing $d{=}3$ spatial
dimensions to $d{=}4{-}\eps$ dimensions.  When $\eps$ is small, one
can systematically remedy the problems of perturbation theory by using
suitable renormalization-group (RG) improved perturbation theory.
Computing to some order in RG-improved perturbation theory corresponds
to computing to some order in $\eps$.  At the end of the day, one
sets $\eps{\to}1$ in the resulting truncated series.
In some cases this is known to give quite good results.%
\footnote{
  A slightly more detailed review for particle theorists, in the context
  of the electroweak phase transition, may be found in the introduction of
  ref.~\cite{arnold&yaffe1}.
}

\begin {figure}
   \begin {center}
        
	\leavevmode
        \figrule
        \epsfbox [65 115 580 670] {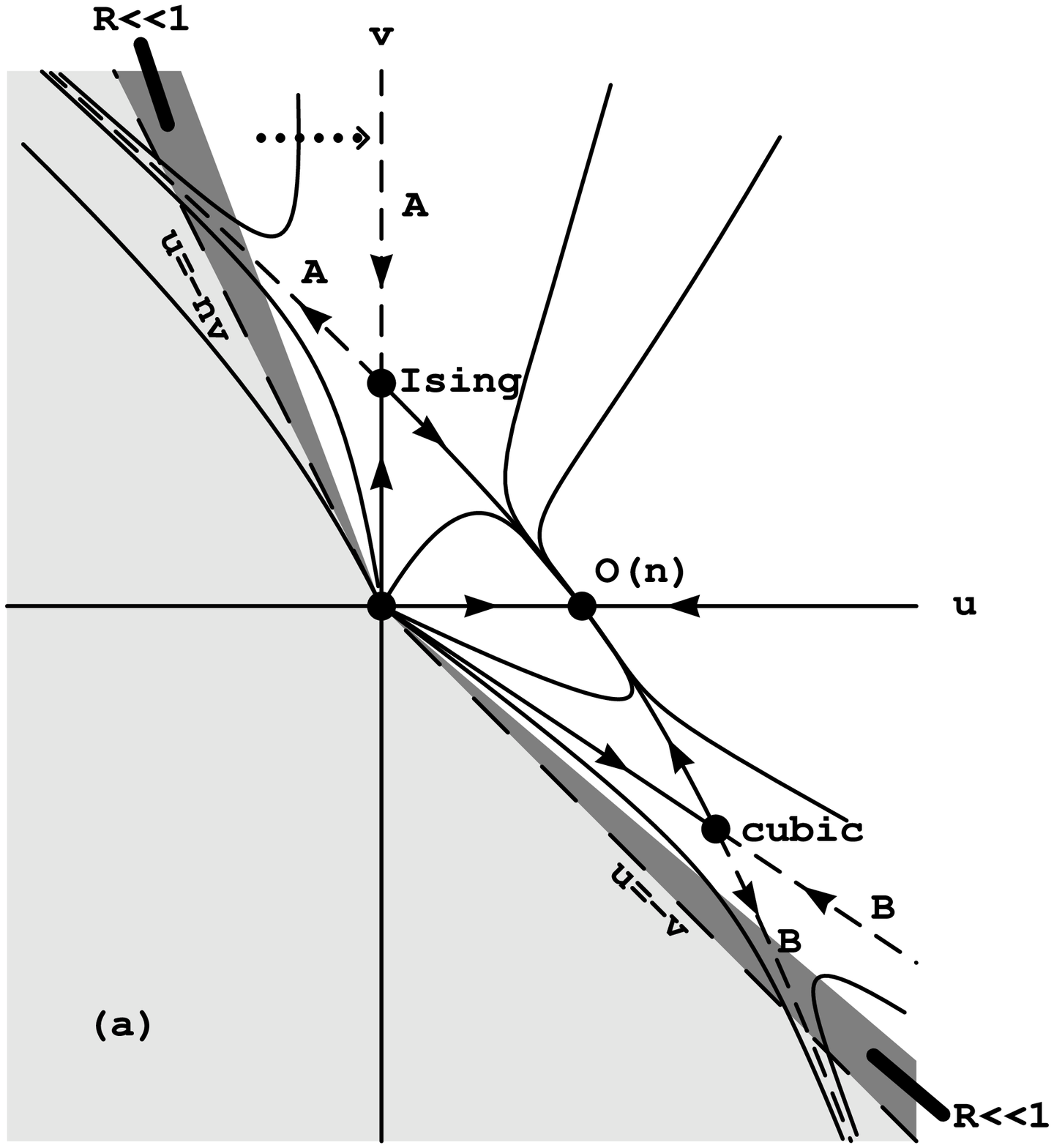}
        \figrule
	\hspace*{8pt}
        
        \figrule
        \epsfbox [65 115 580 670] {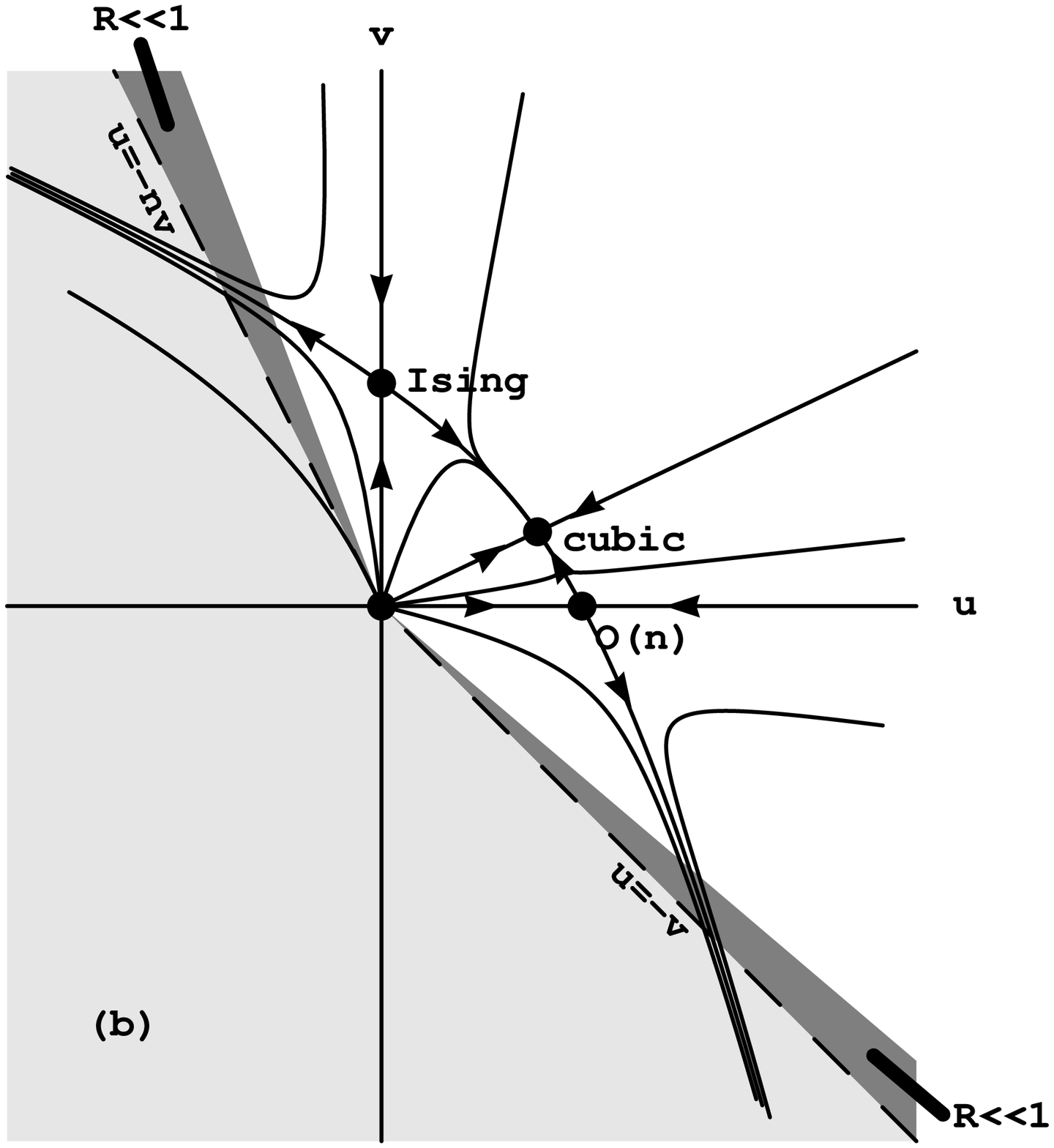}
        \figrule
   \end {center}
   \caption
       {%
	\label {fig:rgflow}
        Renormalization group flow into the infrared of $(u,v)$ for
	small $\eps$ and (a) $n<\nc=4+O(\eps)$, or (b) $n>\nc$.
	The lightly shaded region is the domain in which the
	tree-level potential (\protect\ref{eq:action0}) is unbounded below.
	The more heavily shaded ``wedges'' show the regions in which
	perturbation theory is reliable.
	In the left-hand figure (a),
	the special trajectories labeled A and B will be a focus of
	attention.
	These are the limiting trajectories for theories which
	approach the Ising or cubic fixed points arbitrarily closely
	before flowing off toward the classical instability line.
	Such theories have arbitrarily weak first order phase transitions.
       }%
\end {figure}

Fig.~\ref{fig:rgflow} shows the renormalization group flow, for small $\eps$,
of the dimensionless couplings $u$ and $v$ of the cubic anisotropy model as
one moves to longer distance scales.
The lightly shaded region,
delimited by the lines $u=-n v$ and $u=-v$, designates the range of couplings
where the tree-level potential of (\ref{eq:action0}) is unbounded below.  For
$v=0$, the cubic anisotropy model reduces to an \On\ model, which has a
second-order phase transition associated with the infrared fixed point marked
``\On''.  For $u=0$, the model (\ref{eq:action0}) reduces to $n$ uncoupled
copies of a basic quartic scalar field theory,
which is in the same universality class as
the Ising model.  The associated fixed point is marked ``Ising'' in
fig.~\ref{fig:rgflow}.
Besides the Gaussian fixed point at $u{=}v{=}0$, there is another fixed point
known as the cubic fixed point.  All of these fixed points occur at couplings
of $O(\eps)$, and so the couplings may be treated as small and perturbative
when $\eps\ll 1$.

The stability or instability of these fixed points depends on
the number $n$ of scalar fields.  For $n < \nc = 4 + O(\eps)$
(which encompasses our main case of interest, $n{=}2$), the \On\
fixed point is infrared stable,%
\footnote{
  Except, of course, with respect to the parameter $t$
  ({\it i.e., temperature}), which is not shown
  in the figure and which has to be fine-tuned to reach the phase transition.
}
as shown in fig.~\ref{fig:rgflow}(a),
while the Ising and cubic fixed points have
an unstable direction and correspond to tricritical points.%
\footnote{
   Further expansion in $\eps$ yields
   $\nc = 4 - 2\eps + 2.588\eps^2 + O(\eps^3)$
   \cite{kettley}.
}
A theory whose
couplings lie between the lines running from the origin to the Ising and
cubic fixed points, respectively, will flow at large distances to the
\On\ fixed point and so will have a second-order transition with
\On\ symmetric critical behavior.
A theory with couplings outside of this region
(and not on the critical lines bounding this region) 
does not flow to any weakly coupled infrared-stable fixed point,
and so might be expected to have a first-order phase transition.
This is indeed the case.  As discussed by Rudnick
\cite{rudnick}, and
as we shall review below, the criteria $R \ll 1$ for
the success of perturbation theory (known as the Ginsburg criteria) is
satisfied for couplings very close to the region of tree-level instability,
designated by the heavy shaded regions in Fig.~\ref{fig:rgflow}.
In these $R \ll 1$ regions, one indeed finds that perturbation theory
reliably predicts a first-order transition.
The renormalization group flow then shows that
any theory whose couplings are outside the boundary of the
basin of attraction of the \On\ fixed point
is equivalent to a theory with couplings having $R \ll 1$,
and so will have a first-order transition.
The lines from the origin through the Ising and cubic tricritical points
are therefore boundaries separating theories with first- and second-order
transitions.

The case of $n > \nc$ is shown in fig.~\ref{fig:rgflow}(b).
The \On\ and cubic fixed points exchange roles relative
to the $n < \nc$ case.

Now consider a sequence of theories with first-order transitions,
such as those indicated by the dotted line near the top of fig.~\ref{fig:rgflow}a,
which approach the Ising line $u{=}0$.
The correlation length at the transition can then be
made arbitrarily large, since for $u{=}0$ it is infinite.
The renormalization group trajectory approaches the dashed line in
fig.~\ref{fig:rgflow}a labeled trajectory A.
This limiting trajectory first flows into the Ising fixed point along
the line $u{=}0$, and then flows away from the Ising fixed point in
the unstable direction toward the region of classical instability.
A similar limit for the tricritical behavior of the cubic fixed point
gives trajectory B.
This is the limit we will take to obtain arbitrarily weak
first-order transitions in the cubic anisotropy model.
For $n=2$, trajectories A and B are equivalent because a
redefinition of $\vec\phi$ by a $45^\circ$ internal rotation,
\begin {equation}
   (\phi_1,\phi_2) \to {1\over\sqrt2} (\phi_1+\phi_2, \phi_1-\phi_2) \,,
\end {equation}
leaves the Lagrangian (\ref{eq:action0}) in the same form
but with
\begin {equation}
   (u, v) \to (u + {\textstyle{3\over2}} v, -v) \,.
\label{uv map}
\end {equation}
This means that theories below the $u$ axis in fig.~\ref{fig:rgflow}(a) are,
for $n{=}2$, related by the mapping (\ref {uv map}) to theories above the axis.

We will therefore compute properties of the transition,
for small $\eps$,
on trajectory B.  This is easiest to do by following the trajectory into
the perturbative region $R \ll 1$.  Throughout this paper, we focus
on the flow away from the cubic fixed point.  For $n{=}2$, the final results
for physical quantities must be the same as for flow from the Ising fixed
point.  For other $n$, the flow from the Ising fixed point could be
analyzed similarly, but we have not bothered to do so.

In the next section, we fix notations and renormalization scheme
conventions.  In section~\ref{sec:LO}, we review the leading-order
analysis of the susceptibility ratio $\chi_+/\chi_-$,
which was originally carried
out by Rudnick \cite{rudnick}.  The most straightforward derivation uses a
calculation of the one-loop effective potential and the explicit
one-loop RG equations.  After reviewing this calculation, we show
that explicit knowledge of the one-loop potential and RG equations
was not actually necessary.  In section~\ref{sec:NLO} we extend the
calculation to next-to-leading order.  This calculation does require
the explicit one-loop potential and RG equations, but we show that
explicit knowledge of the two-loop corrections (which would be used
in the most straightforward derivation) is not required.
Section~\ref{sec:NLO xi} computes the next-to-leading order result
for the correlation length ratio $\xi_+^2/\xi_-^2$.
Then, in section \ref{sec:NNLO}, we finally
extend our calculation of $\chi_+/\chi_-$ to next-to-next-to-leading order,
which requires a non-trivial ring-diagram resummation of perturbation theory,
the explicit two-loop potential, and the explicit two-loop RG equations.
Finally, section~\ref{sec:discussion} summarizes our results.
A review of the leading-order result for the flow of the couplings
$(u,v)$, originally derived by Rudnick \cite{rudnick},
as well as details of the two-loop potential, are left to appendices.


\section {Notation and Conventions}

We will use
dimensional regularization
for loop calculations in $d = 4{-}\eps$ dimensions and a
renormalization scheme closely related to
modified minimal subtraction (\MSbar).
Specifically, the bare Lagrangian is
\begin {eqnarray}
   {\cal L}_\bare &=& \half Z_\phi^2 \, |\partial\vec\phi|^2
   + V_\bare(\vec\phi)
\\
\noalign {\hbox {with\footnotemark}}
   V_\bare(\vec\phi) &=&
    \bigl(\normu\bigr)^{-1} \Lambda
    + {\textstyle {1\over2}} Z_{m^2} Z_\phi^2 m^2 |\vec\phi|^2 
    + {\textstyle {1\over4!}} \normu Z_\phi^4 \left[
        Z_u u |\vec\phi|^4 + Z_v v \sum_i \phi_i^4 \right] \,,
\\
\noalign {\hbox {and}}
     \norm &=& (4\pi)^{d/2} \,\Gamma\!\left(\textstyle {d\over2} {-} 1\right) ,
\\
\noalign {\hbox {and where all renormalization constants have the form}}
   Z_i &=& 1 + {z_{i1}(u,v)\over\eps} + {z_{i2}(u,v)\over\eps^2} + \cdots \,.
\end {eqnarray}
\footnotetext
{%
   For $n=2$, our couplings $(u,v)$ are related to Rudnick's \cite{rudnick}
   choice of couplings, call them $(u_{\rm R}, v_{\rm R})$, by $u_{\rm R} =
   (u+v)/12$ and $v_{\rm R} = -v/6$.
}%
Note that we have
rescaled our couplings and $\Lambda$ by an additional factor
of $(4\pi)^2$ compared to the typical convention in particle theory.
The additive constant
$\Lambda$ is irrelevant to the calculation of the susceptibility
or correlation length ratios and may be ignored, but it will be important
for the specific heat ratio computed in ref.~\cite{specific-heat}.
Note that we have relabeled the parameter $t$ as $m^2$, which is the
typical notation used in particle theory.  But it should be kept in mind
that variation of $m^2$ really represents variation of temperature in
the underlying physical problems of interest.%

The susceptibility $\chi$ is defined by adding a linear term $h\phi$ to
the Lagrangian and defining
\begin {equation}
   \chi \equiv \lim_{h \to 0} { d\langle \phi\rangle \over dh } \,.
\end {equation}
The normalization and regularization of the $h\phi$ term is not important
because we will ultimately only be interested in the ratio $\chi_+/\chi_-$,
where it cancels out.


\section {Review of leading-order analysis}
\label{sec:LO}

   Two things are needed to analyze the transition: (a) the location
of some point $(u,v)$ along the portion of the trajectory within
the perturbative regime,
and (b) a perturbative analysis at that point.
Technically, it is easiest to choose the point where the trajectory intersects
the line $u=-nv$ or $u=-v$, where the {\it tree-level\/} potential first
becomes unstable.  (The full, effective potential remains
stable, as it must, since it is invariant under changes
of renormalization scale.)
We shall call this point $(u_*, v_*)$.

The value of $(u_*,v_*)$ turns out
not to affect the leading-order calculation of $\chi_+/\chi_-$;
so we shall proceed for the moment without it.
(The value of the couplings does affect the leading-order
results of other ratios, such as the specific heat ratio $C_+/C_-$
\cite{rudnick,specific-heat}.)

At tree level the transition appears second-order.  For $m^2>0$,
the minimum of the tree-level potential is at $\vec\phi{=}0$.
For $m^2<0$, it is along an ``edge'',
$\vec\phi \propto (1,0,0,\cdots)$, if $v<0$,
and along a ``diagonal'', $\vec\phi \propto (1,1,1,\cdots)$,
if $v>0$.
Rather than discussing the entire structure of the effective
potential $V(\vec\phi)$, it will generally be sufficient simply to
consider its behavior in the relevant direction.  For flow from the
cubic fixed point ($v<0$), attention can be restricted to an edge.
The tree-level potential $V_0$ then becomes
\begin {eqnarray}
   V_0(\vec\phi) \to V_0(\phi)
   &=& {1\over2} m^2 \phi^2 + {1\over 4!} \normu (u+v) \phi^4 \,,
\label{eq:vtree}
\\
\noalign {\hbox {where}}
   \vec\phi &=& (\phi,0,0,\cdots) \,.
\end {eqnarray}
At the instability line, $u{=}{-}v$
and hence the quartic interaction term disappears along the edge.

To see the first-order
nature of the transition, one must consider the effect of
the first loop correction.
This is just the Coleman-Weinberg effect \cite{coleman&weinberg}.
For the sake of definiteness, and because we need the results
later on, we shall go through the explicit calculation of the
one-loop corrections to the effective potential.
However, after the fact, we shall show that an explicit calculation
was actually unnecessary for computing $\chi_+/\chi_-$ at leading order.


\subsection{The one-loop potential}
\label{sec:oneloop}

The one-loop contribution
to the effective potential is
\begin {equation}
   V_1(\vec\phi) = {\cal I}(\ma^2) + (n{-}1) \, {\cal I}(\mb^2)
      ~~+~~  (\hbox{counter-terms}) \,,
\label{eq:V1 general}
\end {equation}
where $\ma^2$ and $\mb^2$ are the eigenvalues of the curvature of
the tree-level potential $V_0(\vec\phi)$ evaluated at $\vec\phi$,
and the one-loop integral ${\cal I}(z)$ is
\begin {equation}
   {\cal I}(z)   = \half \, \tr \ln (-\partial^2+z)
                 = - \half \, (4\pi)^{-d/2} \,
                       \Gamma\left(-\textstyle{d\over2}\right) z^{d/2} \,.
\end {equation}
Along an edge, we have
\begin{mathletters}
\label{eq:mab defs}
\begin {eqnarray}
   \ma^2 &=& m^2 + {\textstyle {1\over2}} \normu (u+v) \, \phi^2 \,,
\label{eq:ma def}
\\
   \mb^2 &=& m^2 + {\textstyle {1\over6}} \normu u \, \phi^2 \,,
\label{eq:mb def}
\end {eqnarray}
\end{mathletters}
which are the curvatures parallel and orthogonal to the edge, respectively.

Now take $u = -v$ to fix ourselves on the tree-level instability line.
It is notationally convenient to express the potential in terms
of \begin {equation}
   M^2 \equiv {\textstyle {1\over6}} \mu^\eps \norm u \, \phi^2,
\label{eq:vnorm}
\end{equation}
and one finds
   \begin {eqnarray}
      \normu V_0(\phi) &=& \Lambda + 3 u^{-1} m^2 M^2 \,,
   \\
      \normu V_1(\phi) &=&
           \left[ {m^4\over2\eps} 
               - {2\pi\mu^\eps \, m^d \over d(d{-}2) \sin(\pi\eps/2)} \right]
         + (n{-}1) \left[ {(m^2{+}M^2)^2\over2\eps} 
         - {2\pi\mu^\eps \, (m^2+M^2)^{d/2} \over d(d{-}2) \sin(\pi\eps/2)}
                     \right] .
   \label {V1 d dim}
   \end {eqnarray}
For the moment, we're only working to leading-order in $\eps$, so we can take
the limit $\eps{\to}0$ to find:%
\footnote{
   Note that leading order in this context doesn't simply mean the
   tree-level potential.
   We are interested in the leading-order results for quantities
   describing the first-order nature of the transition.  But the tree-level
   potential by itself does not describe a first-order transition.
}
   \begin {eqnarray}
      \normu (V_0 + V_1)
      = \Lambda &+& 3 u^{-1} m^2 M^2
      + {1\over4} \, m^4 \left[\,
          \ln\!\left(m^2\over\mu^2\right) - {3\over2} \right]
   \nonumber\\
      &+& {(n{-}1)\over4} \, (m^2+M^2)^2 \left[\,
          \ln\!\left(m^2{+}M^2\over\mu^2\right) - {3\over2} \right]
      + O(\eps) \,.
   \label {V1 with m}
   \end {eqnarray}
As one varies $m^2$,
this potential describes a first-order transition which
occurs at an $m^2{\not=}0$.%
\footnote{%
  By transition, we mean the point where the two ground states are
  degenerate.  In physical applications, this may not be the
  point of direct physical relevance if there is significant super-cooling.
}
If it weren't for the one-loop corrections, the transition would be
second-order and occur at $m^2{=}0$.  Since couplings are $O(\eps)$, one
then expects that $m^2$ at the transition is small if
$\eps$ is small.  We shall indeed see {\it a posteriori} that
\begin {equation}
    {m^2\over M^2} \sim O(\eps)
\end {equation}
when the order parameter $\phi$ is the same order of magnitude as its
value in the asymmetric phase.
So in this range of $\phi$
we can drop $m$ compared to $M$ and find
\begin {equation}
   \normu (V_0 + V_1)
   = \Lambda + 3 u^{-1} m^2 M^2
   + M^4 \left[
        \vconsta \ln\!\left(M^2\over\mu^2\right) + \vconstb \right]
   + O(\eps M^4, m^2 M^2, m^4) \,,
\label {eq:V1loop approx}
\end {equation}
where
\begin {equation}
   \vconsta = {n{-}1\over4} \,,
   \qquad
   \vconstb = -{3\over2} \, \vconsta \,.
\label{eq:vconst values}
\end {equation}
The above approximation to the potential has two degenerate minima when
\begin {equation}
   m^2 = m_1^2 \equiv {\vconsta \, u \, \mu^2 \over 3}
     \exp\left( -1 -{\vconstb\over\vconsta} \right)
     = {n{-}1 \over 12} \, u \, \mu^2 e^{1/2}
     \,,
\label{eq:m1}
\end {equation}
and the value of $M$ at the asymmetric minima is
\begin {equation}
   M^2 = M_1^2 \equiv \mu^2
     \exp\left( -1 -{\vconstb\over\vconsta} \right)
     = \mu^2 e^{1/2}
     \,.
\label{eq:M1}
\end {equation}
As promised, $m_1^2/M_1^2 \sim O(u) \sim O(\eps)$.
The curvatures at the origin and at the asymmetric minima are%
\footnote{
   The proportionality relationship reflects the fact that the derivatives on
   the right-hand side are with respect to $\phi(\mu)$---the renormalized
   field at the scale where $u(\mu)=-v(\mu)$.  However, this $\phi$ is
   proportional to the bare $\phi$, and thus the ratio (a) does not depend on
   the proportionality constant, and (b) is insensitive to short-distance
   physics.
}
\begin {eqnarray}
    1/\chi_+
&\propto&
    \left.{\partial^2 V(\phi) \over \partial \phi^2}\right|_{\phi=0}
     = {\textstyle{1\over6}} \normu u V''(0)~~~\>
     = m^2 \, [1 + O(\eps)] \,,
\\
   1/\chi_-
&\propto&
    \left.{\partial^2 V(\phi) \over \partial \phi^2}\right|_{\phi=\phi_1}
     = {\textstyle{1\over6}} \normu u V''(M_1)
     = 2 m^2 \, [1 + O(\eps)]
\end {eqnarray}
(where $V'(M) \equiv \partial V(M) / \partial M$, {\em etc.}).
Thus,
\begin {equation}
   {\chi_+\over\chi_-} = 2 + O(\eps) \,.
\label {chi ratio 1loop}
\end {equation}


\subsection{Avoiding one-loop details}
\label{sec:avoidoneloop}

  In the preceding derivation, the final result for $\chi_+/\chi_-$
did not, in fact, depend on the values (\ref{eq:vconst values})
of the constants $\vconsta$ and $\vconstb$.
So we could have
arrived at the same result simply knowing the form (\ref{eq:V1loop approx})
of the one-loop potential in the limits $\eps{\to}0$, $u{=}{-}v$,
and $m^2/M^2 \sim O(\eps)$.
But this form just follows from (a) the existence of the renormalization group,
which produces a single power of $\ln\mu$ at one-loop order, and
(b) the fact that $M$ is the only relevant dimensionful parameter if
$m$ is negligible.  The logarithm must therefore be $\ln(M/\mu)$ in our
approximation.
If we had understood {\it a priori} the
independence of the result on $\vconsta$ and $\vconstb$, we could
have avoided doing the explicit one-loop calculation.
Generalizations of the following arguments
will later save us from the need for two-loop calculations at
next-to-leading order,
and three-loop calculations at next-to-next-to-leading order.

   The independence from $\vconstb$ may be understood as follows.
The only parameters that final results can depend on are
the dimensionless couplings $(u_*,v_*)$ and the corresponding
renormalization scale $\mu$.  Other parameters, such as $m^2$
and the scalar expectation $M$ have been solved for and eliminated by
requiring that we be at the transition and in one or the other phase.
If our final result is a dimensionless ratio, such as $\chi_+/\chi_-$
or $C_+/C_-$,
it must then be independent of $\mu$ and can depend only on
$(u_*,v_*)$.  So the answer can't change even if we arbitrarily
change $\mu$ in (\ref{eq:V1loop approx}) while holding
$(u_*,v_*)$ fixed.  (This is different from a simple statement of
RG invariance, which would involve changing $(u,v)$ in a compensating
manner when changing $\mu$.)  Such a change in $\mu$ (at lowest order)
is equivalent to varying $\vconstb$,
and so dimensionless ratios do not depend on $\vconstb$.

   The coefficient $\vconsta$ of the logarithm is determined by the
one-loop renormalization group.
However, as noted above, it is not needed for $\chi_+/\chi_-$.
The reason is that, at leading order, the result (\ref{chi ratio 1loop}) is
independent of the couplings $(u_*,v_*)$.  So the result would be the
same if we redefined $(u,v)$ by $(u,v) \to (xu,xv)$ for some constant $x$.
This redefinition does not take us off the line $u{=}{-}v$, but it does
change the coefficients of the
one-loop $\beta$-functions 
and therefore changes $\vconsta$.
The leading-order result for $\chi_+/\chi_-$ must
therefore be independent of $\vconsta$.
This simplification doesn't occur for $C_+/C_-$, which turns out to be
proportional to $u_*$ at leading order \cite{rudnick,specific-heat}.

    As a prelude to our later higher-order analysis, it will be useful
to sketch the renormalization group determination of $\vconsta$.
The one-loop RG equation for the effective potential in
the cubic anisotropy model is
\begin {equation}
   \left(
       \mu {\partial \over \partial \mu}
     + \beta_u {\partial \over \partial u}
     + \beta_v {\partial \over \partial v}
     + \beta_{m^2} \, m^2 {\partial \over \partial {m^2}}
     + \gamma_\phi \, \phi {\partial \over \partial \phi}
     + \beta_\Lambda {\partial \over \partial \Lambda}
   \right)
   V = 0 \,,
\label{eq:RG1}
\end {equation}
where%
\footnote
    {%
    The trivial $\epsilon$ dependence in the beta functions
    (\ref {eq:d dim beta}) is, of course, a standard
    feature of minimal subtraction renormalization schemes.
    }
\begin {mathletters}%
\label{eq:d dim beta}%
\begin {eqnarray}
   \beta_u(u,v,\eps) &=& -\eps u + \bar\beta_u(u,v) \,,
\\
   \beta_v(u,v,\eps) &=& -\eps v + \bar\beta_v(u,v) \,,
\end{eqnarray}
\end {mathletters}%
with
\begin {mathletters}%
\begin{eqnarray}
   \bar \beta_u(u,v) &=& \betaone_u + O(u^3,v^3) \,,
   \qquad
   \betaone_u = u\left({\textstyle {1\over3}} (n{+}8) \, u + 2 v\right) \,,
\\
   \bar \beta_u(u,v) &=& \betaone_v + O(u^3,v^3) \,,
   \qquad
   \betaone_v = v \left(4u + 3v \right) \,,
\\
   \beta_{m^2}(u,v) &=& \betaone_{m^2} + O(u^2,v^2) \,,
   \qquad
   \betaone_{m^2} = {\textstyle {1\over3}} (n{+}2) \, u + v \,,
\\
   \gamma_\phi(u,v) &=& O(u^2,v^2) \,.
\end {eqnarray}%
\label{eq:beta1}%
\end {mathletters}%
For the susceptibility, we are only interested in the $\phi$ dependence
of the potential, and the running of $\Lambda$ will be irrelevant.
As it turns out, we will also not need $\beta_{m^2}$ or
$\gamma_\phi$ below.

The renormalization group flow does not map the line $u{+}v=0$
onto inself.
Consequently,
to apply the RG equation, one must retain the $v$
dependence of the tree-level potential (\ref{eq:vtree})
rather than specializing from the outset to $u{=}{-}v$.
Working at $\eps{=}0$, one easily finds that
the RG equation is satisfied by
\begin {eqnarray}
   V_0 + V_1
   &=&
       V_0 - \left\{
       {\textstyle {1\over2}} \, \betaone_{m^2} \, m^2\phi^2
      + {\textstyle {1\over4!}} \,
      \norm[\betaone_u+\betaone_v] \, \phi^4
             \right\} \ln\mu
\nonumber\\ && \qquad\qquad
     {} + (\hbox{$\mu$-independent})
     + O(\eps^2 V) \,.
\end {eqnarray}
%
Taking $u{=}{-}v$
(and neglecting $m^2$ relative to $M^2$) then gives
\begin {equation}
   \vconsta = {3\over4u^2} \left(\betaone_u + \betaone_v\right)
	       \biggl|_{u{=}{-}v} \,,
\end {equation}
which agrees with (\ref{eq:vconst values}).
In later sections,
this same RG method will be used to determine
$\mu$-dependent terms in the two- and three-loop contributions
to the effective potential.



\subsection{Scale hierarchies and subtleties at higher orders}
\label{hierarchy section}

As we shall see, the preceding tricks for simplifying calculations
will generalize to higher orders in $\eps$ as well.
Leading-order results for generic ratios require only explicit
knowledge of the tree-level potential and the one-loop renormalization
group, and $\chi_+/\chi_-$ doesn't even need the latter.
Next-to-leading order calculations generically require
the explicit one-loop potential and the two-loop renormalization group,
although $\chi_+/\chi_-$ needs only the one-loop renormalization group.
Next-to-next-to-leading order generically requires the two-loop potential
and the three-loop renormalization group, and so forth.

\begin {figure}
\vbox
    {%
    \begin {center}
	
	\figrule
	\epsfbox [85 340 485 460] {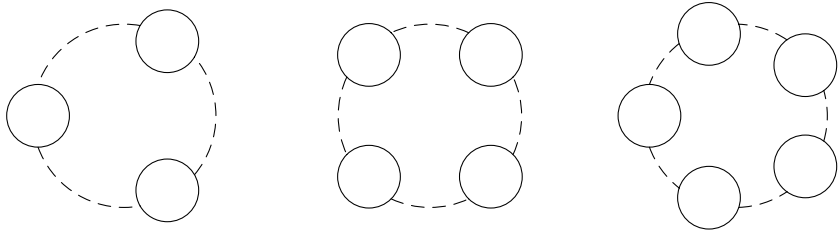}
	\figrule
    \end {center}
    \caption
	{%
	\label {fig:rings}
        A sequence of diagrams of the same order in $\eps$ in the asymmetric
        phase.  Solid lines represent heavy ($\mb$) degrees of freedom;
        light lines represent the light ($\ma$) degrees of freedom.
	}%
    }%
\end {figure}

There is a subtlety, however, in the simplistic assumption that each successive
order in $\eps$ requires exactly one more order in the loop expansion of
the effective potential.  The source of this subtlety is the ratio of
scales $m^2/M^2 \sim O(\eps)$ in the asymmetric phase.
Fig.~\ref{fig:rings} shows
a sequence of diagrams, with arbitrarily many loops (starting at 3 loops),
which are all the same order in $\eps$ in the asymmetric phase.
These diagrams consist of multiple ``ring'' corrections to the light
(mass $m^2$) mode due to interactions with the heavy (mass $M^2$) modes.
The outer loop is dominated by momenta of order $m$, and the
cost of adding an additional ring is
\begin {equation}
      O(u^2 \phi^2 / m^2) = O(u M^2/ m^2) = O(1) \,.
\end {equation}

This particular problem could be handled diagrammatically by resumming
the light propagator to incorporate all one-loop heavy rings.
A more elegant way to think about it is in the language of the
renormalization group.  At distances large compared to $1/M$ in
the asymmetric phase, our scalar theory should be replaced by an
effective theory consisting of only the light degree of freedom.
The mass of the light scalar in the effective theory will be its
original mass $m$ plus corrections from integrating out the
heavy modes.

The free energy in this light effective theory will be of order
$m^4$.  Comparison to
(\ref{eq:V1loop approx})
then reveals that it will only be important
at next-to-next-to-leading order.
If only working to NLO, one can ignore the need for
this resummation in the asymmetric phase.


\section {NLO analysis of $\chi_+/\chi_-$}
\label{sec:NLO}

\subsection{The effective potential: asymmetric phase}
\label{sec:asymtwoloop}

Going one order beyond the previous analysis requires consideration of
(a) two-loop contributions to the effective potential,
and (b) corrections to the $\eps \to 0$ and $m^2/M^2 \to 0$ limits
we took of the one-loop potential.
For the latter, we can simply expand the general one-loop potential
(\ref{V1 d dim}) to the desired order:
\begin {eqnarray}
   \normu (V_0 + V_1)
   &=& \Lambda + 3 u^{-1} m^2 M^2
   + M^4 \left[
        \vconsta \ln\!\left(M^2\over\mu^2\right) + \vconstb \right]
   + \normu \delta V_1
   + O(\eps^2 V)_\asym \,,
\label{eq:v0 plus v1}
\\\noalign {\hbox {with}}
     \normu \, \delta V_1 &=&
     \eps \, M^4 \left[
         -{\textstyle {1\over4}}\vconsta \ln^2 \!\left(M^2\over\mu^2\right)
         -{\textstyle {1\over2}}\vconstb \ln \!\left(M^2\over\mu^2\right)
         + \vconstc
       \right]
\nonumber\\ &+&
     m^2 M^2 \left[\,
         2 \vconsta \ln\!\left(M^2\over\mu^2\right) - 2 \vconsta  \right],
  \label {eq:delta V1}
\\\noalign {\hbox {and}}
   \vconstc &=& -\left({7\over8} + {\pi^2\over12}\right) \vconsta \,.
\label{eq:vconstc value}
\end {eqnarray}
The subscript ``$\asym$'' is a reminder of the assumption
$m^2/M^2 \sim O(\eps)$ in the error estimate, which is valid only
in the asymmetric phase.

For the two-loop contribution, we may take $\eps{=}0$ and ignore
$m^2$ altogether.
The renormalization group requires the contribution to have the form
\begin{equation}
   \normu V_2 = u M^4 \left[\,
        \vconst_{22}\ln^2\!\left(M^2\over\mu^2\right)
      + \vconst_{21}\ln\!\left(M^2\over\mu^2\right)
      + \vconst_{20} \right]
      + O(\eps^2 V)_\asym \,.
\label{eq:V2 form}
\end{equation}
As in the leading-order calculation, we will not actually need to compute
all three parameters.

The first simplification is to note that rescaling $\mu$ by
\begin {equation}
   \mu \to \mu \, (1 + xu)
\end {equation}
in $V_0+V_1+V_2$, while holding all couplings fixed, changes $\vconst_{20}$ at
this order but nothing else.  Therefore dimensionless ratios cannot
depend on $\vconst_{20}$ at NLO.  Similarly, a change such as
\begin {equation}
   \mu \to \mu \, (1 + x\eps)
\end {equation}
would change $\vconstc$ and nothing else, so we never actually needed
its value (\ref{eq:vconstc value}) for a NLO calculation.

We could determine all the other constants in (\ref{eq:V2 form}) by
requiring the potential to satisfy the RG equation at two loops.
However, analogous to what happened at leading order, we will not
need all of these coefficients for $\chi_+/\chi_-$.  It is sufficient
to apply the RG equation at {\it one\/} loop order, as given by
(\ref{eq:RG1}--\ref{eq:beta1}).
However, we do need the one-loop potential for
general $(u,v)$ without the restriction $u{=}{-}v$.  Returning to
(\ref{eq:V1 general}) and (\ref{eq:mab defs}), one easily finds
   \begin {eqnarray}
      \norm \mu^\eps \, (V_0 + V_1)
      &=&
	  \Lambda
	+ {3\over2} \, {u{+}v \over u^2} \, M^4
        + {9\over4} \, {(u{+}v)^2 \over u^2} \, M^4
	    \left[
             \ln \!\left(3 \, (u{+}v) \, M^2\over u \, \mu^2\right) - {3\over2}
	     \right]
   \nonumber\\&& \qquad\qquad\qquad\qquad {}
        + {(n{-}1)\over4} \, M^4
	    \left[
             \ln \!\left( M^2\over\mu^2\right) - {3\over2} \right]
        + O(\eps^2 V)_\asym \,.
   \label{eq:1loop w u+v}
   \end {eqnarray}
Applying the one-loop RG equation,%
\footnote
    {%
    When applying the renormalization group equation,
    it is helpful to note that $M^2$ is multiplicatively
    renormalized,
    $
	\mu (\partial M^2 / \partial \mu) = M^2 \, \beta_{M^2}
    $,
    with
    $
	\beta_{M^2} \equiv \bar\beta_u / u + 2 \gamma_\phi
	= \beta^{(1)}_u + O(u^2)
    $.
    }
and then setting $u{=}{-}v$, determines
\begin {equation}
   \vconst_{22} = {n{+}2\over6} \vconsta \,.
\label{eq:vconst22}
\end {equation}
It is also worth noting that it was unnecessary to compute explicitly the
$O(\eps)$ correction to $V_1$ given by the first term of
(\ref{eq:delta V1}),
because the coefficients of the logs in that correction
are determined by the RG equation as well, arising from the explicit $\eps$
in (\ref{eq:d dim beta}).  This observation will substantially simplify
the analogous calculation when we later proceed to NNL order.

To find the asymmetric phase susceptibility $\chi_-$, we now need at
next-to-leading order both
the value of $m^2$ at the transition and of $M^2$ in the asymmetric
phase.  Perturb around the one-loop solutions (\ref{eq:m1}) and
(\ref{eq:M1}) by writing
\begin {equation}
   V = V_{(1)} + \delta V \,,
   \qquad
   m^2 = m_1^2 + \delta m^2 \,,
   \qquad
   M = M_1 + \delta M \,,
\end{equation}
where $V_{(1)}$ is the one-loop $\eps{=}0$
approximation (\ref{eq:V1loop approx}) to the effective potential and
\begin {equation}
   \delta V = \delta V_1 + V_2
\end {equation}
as parameterized in (\ref{eq:delta V1}) and (\ref{eq:V2 form}).
By linearizing the equations $V(M){=}V(0)$ and $V'(M) {=} 0$
that determine the asymmetric phase expectation $M$ at the
transition point, one finds
\begin {eqnarray}
   \delta m^2 &=&
       - \norm \mu^\eps \, {u \over 3}
	   \left.
             \left( \delta V(M) - \delta V(0)
                  \over M^2 \right)
             \right|_{M_1,m_1}
       + O(\eps^2 m^2) \,,
\\
\noalign {\hbox {and}}
   \delta M &=&
       - {M^2\over V_{(1)}''} \,
      {\partial \over \partial M} \! \left. \left( \delta V\over M^2 \right)
       \right|_{M_1,m_1}
       + O(\eps^2 M) \,.
\end {eqnarray}
The fractional shift in $1/\chi_-$ is
\begin {equation}
   \left({d^2 V_{(1)} \over d M^2}\right)^{-1}
   \delta {d^2 V\over d M^2} =
     \left.
     \left(
       {\partial^2 \, \delta V \over \partial M^2}
       + {6 \, \delta m^2 \over \norm \mu^\eps u} 
       + \delta M \, {\partial^3 V_{(1)} \over \partial M^3}
    \right)
     \right|_{M_1, m_1} .
\end {equation}
Putting in the explicit form for $\delta V$, we obtain
  \begin {eqnarray}
     m^2 &=& m_1^2 \left[
       1 + \left(-{5\over16} - {\vconstc\over\vconsta}\right)\eps
         + \left({(n{-}4)\over24} - {1\over2} {\vconst_{21}\over\vconsta}
              - {\vconst_{20}\over\vconsta}\right)u
         + O(\eps^2) \right] \,,
  \\
     M^2 &=& M_1^2 \left[
       1 + \left(-{13\over16} - {\vconstc\over\vconsta}\right)\eps
         + \left(-{(3n{+}2)\over8} - {3\over2} {\vconst_{21}\over\vconsta}
              - {\vconst_{20}\over\vconsta}\right)u
         + O(\eps^2) \right] \,,
  \\
     1/\chi_- &\propto&
     \partial_\phi^2 V(\phi) =
     2 m^2 \left[ 1 - {1\over2}\, \eps
       - {(n{-}13)\over12} \, u
       + O(\eps^2) \right] \,.
  \label{eq:chi- NLO}
  \end {eqnarray}
Note how all the dependence of $\partial_\phi^2 V$ on
$\vconstc$, $\vconst_{21}$, and $\vconst_{20}$
in the result (\ref{eq:chi- NLO})
is hidden in the overall factor of $m^2$.


\subsection{The effective potential: symmetric phase}

When examining the asymmetric phase, we made an expansion in
$m^2/M^2 = O(\eps)$.  For the symmetric phase, where $M{=}0$,
this is not a good approximation.  The symmetric phase is easier, however,
because one needs only one-loop contributions at $\eps{=}0$
for next-to-leading order results.  Consider the one-loop potential
(\ref{V1 with m}).  It gives
\begin {eqnarray}
  1/\chi_+ \propto
  \partial_\phi^2 V(0) &=&
  m^2 \left\{1
    + {(n{-}1)\over6} \, u \left[\ln\!\left( m^2\over\mu^2 \right) - 1\right]
    + O(\eps^2) \right\}
\nonumber\\
  &=&
  m^2 \left\{1 + {(n{-}1)\over6} \, u \left[ \ln\!\left( {(n{-}1)\over12}u \right)
                    - {1\over2} \right]
     + O(\eps^2) \right\} \,.
\end {eqnarray}
Putting this together with the asymmetric phase result gives our NLO ratio
\begin {equation}
   {\chi_+\over\chi_-} = 2 \left[
     1 - {1\over2} \, \eps
     - {(n{-}1)\over6} \, u_* \ln\!\left({(n{-}1)\over12} u_*\right)
       + u_* \right] + O(\eps^2) \,,
\label{eq:NLO ratio n}
\end {equation}
where we now explicitly remind ourselves that $u$ is to be evaluated at
the point $(u_*,v_*)$.  At leading order, we didn't need to know $u_*$
at all for $\chi_+/\chi_-$.
For next-to-leading order, we need to find the leading order value of $u_*$.


\subsection{$(u_*,v_*)$ at leading order}
\label{sec:LOflow}

The one-loop RG equations for the couplings are
\begin {mathletters}
\label{eq:uv RG1}
\begin {eqnarray}
   \mu{\partial u \over \partial \mu} &=& -\eps u + \betaone_u(u,v) \,,
\\
   \mu{\partial v \over \partial \mu} &=& -\eps v + \betaone_v(u,v) \,,
\end {eqnarray}%
\end {mathletters}%
where $\betaone_u$ and $\betaone_v$ are given in (\ref{eq:beta1}).
The explicit solution was found when $n{=}2$
by Rudnick \cite{rudnick} and generalized
to other $n$ by Domany, {\it et al.} \cite{domany}.
We review the derivation in
Appendix~\ref{apndx:ustar}.  The resulting trajectories are given by
\begin {eqnarray}
   v &=& \eps \, R(u/v,c) \,,
\label{trajectories}
\\\noalign {\hbox {where}}
   R(f,c) &\equiv&
      {\lambda^2 \, f^{-3} \over (n\lambda+1)(n\lambda+2)}
      \left[ {(n\lambda+1)\over\lambda} f^2 - 2 f
      + {2\over n}
      - {2c\over n} \left(1 + {f\over\lambda}\right)^{-n\lambda}
      \right] \,,
\label{eq:Rfc}
\\
   \lambda &\equiv& {3\over 4-n} \,,
\label{eq:lambda}
\end {eqnarray}
and each choice of the constant $c$ picks out a different trajectory.
The trajectory that flows away from the cubic fixed point is $c=0$;
the one flowing away from the Ising fixed point is $c=1$.
The values of $(u,v)$ on the tree-level instability line are then
\begin {equation}
   (u_*, v_*) = (u_*, -u_*) \,,
   \quad
   u_* = - \eps R(-1,0) = {3 (n^2+5n+3) \over n (n{+}2) (n{+}8)} \, \eps \,,
\label{eq:LO u n}
\end {equation}
for flow from the cubic fixed point and
\begin {equation}
   (u_*, v_*) = (-{\textstyle{1\over n}} v_*, v_*) ,
   \;\;
   v_* = \eps R(-{\textstyle {1\over n}},1) = {3 n \over (n{+}2) (n{+}8)}
      \left[ \textstyle 3n
	  \left({4\over 3}(1{-}{1\over n})\right)^{-n\lambda} -(7n{+}2) \right]
      \eps ,
\end {equation}
for flow from the Ising fixed point.  For $n{=}2$, these are
\begin {equation}
   \left( \textstyle{51\over80} \eps, -\textstyle{51\over80} \eps \right)
   \quad \hbox{and} \quad
   \left( -\textstyle{51\over160} \eps, \textstyle{51\over80} \eps \right) ,
\label {eq:n=2 ustar}
\end {equation}
respectively, and are related by the mapping (\ref{uv map}).

Our final, next-to-leading order result for $\chi_+/\chi_-$ is given by
(\ref{eq:NLO ratio n}) and (\ref{eq:LO u n}).  For $n{=}2$ this yields
\begin {equation}
   {\chi_+\over\chi_-} = 2 \left[
      1 - {17\over160}\, \eps \ln\!\left(17\eps\over320\right)
       + {11\over80}\, \eps + O(\eps^2) \right] \,.
\label{eq:NLO ratio 2}
\end {equation}
The presence of an $\eps \ln \eps$ term
is a feature that one does not encounter in $\eps$ expansions
for critical exponents of second-order transitions.  It arises here from
the hierarchy of scales $m^2/M^2 \sim \eps$ characterizing the physics of
the asymmetric phase at the transition.


\section {NLO analysis of $\xi_+^2/\xi_-^2$}
\label{sec:NLO xi}

\begin {figure}
\vbox
    {%
    \begin {center}
	
	\figrule
	\epsfbox [135 330 475 480] {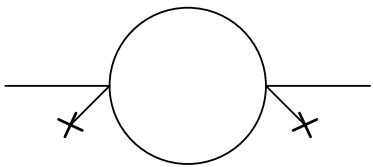}
	\figrule
    \end {center}
    \caption
	{%
	\label {fig:delta pi}
        One loop diagram contributing to momentum-dependence of $\Pi$
        in asymmetric phase.
	}%
    }%
\end {figure}

The correlation length $\xi$ is determined by the location of the pole of the
two-point correlation.  This is the solution $p^2 = -\xi^{-2}$
to
\begin {equation}
   p^2 + m^2 + \Pi(p^2) = 0 \,,
\end {equation}
where $\Pi(p^2)$ is the (one-particle irreducible) self-energy.
Since $m^2{+}\Pi(0)$ is another name for the susceptibility $1/\chi$, we can
write
\begin {equation}
   \xi^{-2} = \chi^{-1} + \left[\Pi(-\xi^{-2}) - \Pi(0)\right] \,.
\end {equation}
As we shall see, this equation can be solved by iteration, treating the
second term, call it $\Delta\Pi$, as small.
Hence,
at leading order
$\xi^{-2}$ is the same as $\chi^{-1}$ and is $O(m^2)$.
Fig.~\ref{fig:delta pi} shows the only
one-loop graph that contributes to the momentum dependence of $\Pi(p^2)$.

In the symmetric phase, the expectation $\langle\phi\rangle$ is zero and
so fig.~\ref{fig:delta pi} vanishes.
The only mass scale is $m$ and so two-loop contributions to
$\Delta\Pi$ would be order $O(u^2 m^2) = O(\eps^2 \xi^{-2})$,
which can be ignored at next-to-leading order.

In the asymmetric phase, the largest correlation length will be that
associated with the degree of freedom $\phi_1$ along the edge,
corresponding to $m_a^2=O(m^2)$ in (\ref{eq:mab defs}).
For $u=-v$, this degree of freedom does not couple to itself, and the
degrees of freedom running around the loop in fig.~\ref{fig:delta pi} are the
heavier ones $\phi_i$ ($i\not=1$) associated with the mass scale
$\mb^2 = O(M^2)$.  The momentum dependence $\Delta\Pi$
of fig.~\ref{fig:delta pi} is
therefore
$O(u\, p^2) = O(u\, m^2) = O(\eps \, \xi^{-2})$.  As advertised, it can be
treated as a perturbation.
Explicit calculation in the $p^2 \ll M^2$ limit gives
\begin {equation}
   \Pi(p^2) - \Pi(0) = {(n{-}1)\over18} \,u \,p \,^2 + O(\eps^2 m^2)_{\asym} \,,
\end {equation}
and so
\begin {eqnarray}
   \xi_-^2 &=& \chi_-
       \left[\, 1 + {(n{-}1)\over18}\, u + O(\eps^2)\right] \,,
\\
   \xi_+^2 &=& \chi_+
       \left[\, 1 + O(\eps^2)\right] \,.
\end {eqnarray}
Combining with the result (\ref{eq:NLO ratio n})
for the susceptibility ratio gives
\begin {equation}
   {\xi_+^2\over\xi_-^2} = 2 \left[
     1 - {1\over2} \eps
     - {(n{-}1)\over6} u_* \ln\!\left({(n{-}1)\over12} u_*\right)
     - {(n{-}19)\over18} u_* \right] + O(\eps^2) \,.
\label{eq:NLO xi ratio n}
\end {equation}
Finally, inserting the $n{=}2$ value (\ref {eq:n=2 ustar}) of $u_*$
yields
\begin {equation}
   {\xi_+^2\over\xi_-^2} = 2 \left[\,
      1 - {17\over160}\, \eps \ln\!\left(17\eps\over320\right)
       + {49\over480}\, \eps + O(\eps^2) \right] \,.
\label{eq:NLO xi ratio 2}
\end {equation}


\section {NNLO analysis of $\chi_+/\chi_-$}
\label{sec:NNLO}

\subsection{The 2-loop renormalization group}
\label{sec:twoloopRG}

Two-loop RG $\beta$-functions for the cubic anisotropy model may be
easily extracted by following standard derivations in one-scalar
models and replacing the overall couplings of each diagram by those
appropriate for our two-scalar model.  One finds
\begin {eqnarray}
   \bar\beta(u,v) &=& \betaone + \betatwo + O(u^4,v^4) \,,
\\
   \gamma(u,v) &=& \gammatwo + O(u^4,v^4) \,,
\\\noalign {\hbox {with}}
   \betatwo_u &=&
       - {(3n+14)\over3} u^3 - {22\over3} u^2 v - {5\over3} u v^2 \,,
\\
   \betatwo_v &=&
       - {(5n+82)\over9} u^2 v - {46\over3} u v^2 - {17\over3} v^3 \,,
\\
   \betatwo_{m^2} &=& - {5\over6} \left[
       {(n{+}2)\over3} u^2 + 2 u v + v^2 \right] ,
\\
   \gammatwo_\phi &=& - {(n{+}2)\over36} u^2 - {1\over6} uv - {1\over12} v^2 \,.
\end {eqnarray}


\subsection{The effective potential: asymmetric phase}

\begin {figure}
\vbox
    {%
    \begin {center}
	
	\figrule
	\epsfbox [100 350 480 445] {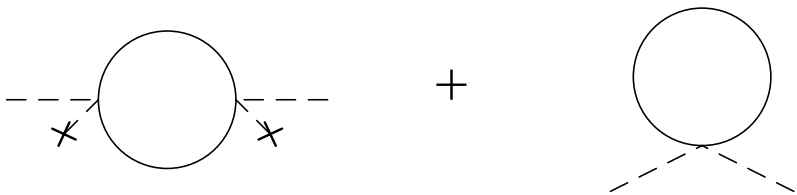}
	\figrule
    \end {center}
    \caption
	{%
	\label {fig:meff}
        One-loop diagram contributing to effective light mass.
	}%
    }%
\end {figure}

As discussed in section~\ref{hierarchy section}, the NNLO analysis of
the effective potential in the asymmetric phase requires separating
the light and heavy modes and doing a resummation of the heavy modes'
effect on the light ones.  The effective mass $\meff$ of the light mode
at distances large compared to $1/M$ can be found by explicitly
computing the diagrams of fig.~\ref{fig:meff}.
More simply, it can be taken from
the curvature of the one-loop potential (\ref{eq:V1loop approx})
near the asymmetric minima (\ref{eq:M1}):
\begin {equation}
   \meff^2(M) = m^2
       + 3 {(u{+}v) \over u} M^2
       + (n{-}1) \, u M^2 \left[ {1\over2} \ln\!\left(M^2\over\mu^2\right)
                           - {1\over6}
			   \right] + O(\eps m^2) \,.
\label{eq:meff}
\end{equation}
The sub-leading $O(\eps m^2)$ corrections to the above relationship
are convention dependent: they depend on exactly how we want to define
$\meff$.  However, such sub-leading corrections to $\meff^2$ are
not relevant at the order of interest.

The effective potential in the asymmetric phase is
\begin {eqnarray}
   V = V_0
          ~&+&~ ( V_1^\heavy
                 + V_1^\light
                 + \delta V_1^\heavy
                 + \delta^2 V_1^\heavy )
\nonumber\\ ~&+&~
	      ( V_2^\heavy
                 + \delta V_2^\heavy )
          ~+~ V_3^\heavy
          ~+~ O(\eps^3 V)_\asym \,,
\label{eq:V asym}
\end {eqnarray}
where $V_0 + V_1^\heavy$ and $\delta V_1^\heavy$ are given
by (\ref{eq:v0 plus v1}) and (\ref{eq:delta V1}).  We have added
the superscript ``$\heavy$'' to indicate that these contributions come
from the heavy modes.  The next correction $\delta^2 V_1^\heavy$ from the
expansion of the second term in the one-loop potential (\ref{V1 d dim}) is
\begin {eqnarray}
   \normu \, \delta^2 V_1^\heavy &=&
   \vconsta \, m^4 \ln\!\left(M^2\over\mu^2\right)
   + \eps \, m^2 M^2 \left[
       - {\textstyle {1\over2}} \,\vconsta \ln^2 \!\left(M^2\over\mu^2\right)
       + \vconsta \ln \!\left(M^2\over\mu^2\right)
       + \vconstcm
     \right]
\nonumber\\ &+&
   \eps^2 M^4 \left[
       {\textstyle {1\over24}} \,\vconsta \ln^3 \!\left(M^2\over\mu^2\right)
       +{\textstyle {1\over8}} \,\vconstb \ln^2 \!\left(M^2\over\mu^2\right)
       -{\textstyle {1\over2}} \,\vconstc \ln \!\left(M^2\over\mu^2\right)
       +\vconstd
     \right] .
\label {eq:delta2 V1}
\end {eqnarray}
This can be justified either by explicit expansion or by application of
the renormalization group.  The new constant $\vconstd$ will
be irrelevant because it can be absorbed into a redefinition of
$\mu$.  The new constant $\vconstcm$ may be extracted from explicit expansion
of (\ref{V1 with m}):
\begin {equation}
   \vconstcm = -\left(1 + {\pi^2\over6}\right) \vconsta \,.
\label {eq:vconstcm}
\end {equation}

\begin {figure}
\vbox
    {%
    \begin {center}
	\figrule
	\epsfbox {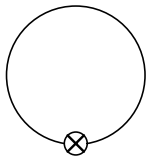}%
	\figrule
    \end {center}
    \caption
	{%
	\label {fig:counterterm}
        Next-to-leading order contribution involving
	the effective mass counterterm $(m^2 {-} \meff^2)$.
	}%
    }%
\end {figure}

The last piece of the one-loop potential we need is the contribution
from the light modes, corresponding to the third term of (\ref{V1 with m}). 
However, this contribution must be computed with the correct effective mass
(\ref{eq:meff}) so that
\begin{equation}
   \normu V_1^\light = 
      {1\over4} \, \meff^4 \left[
        \ln\!\left(\meff^2\over\mu^2\right) - {3\over2} \right] \,.
\label{eq:V1 light}
\end {equation}
As usual, this mass resummation is most easily accomplished by rewriting
the light mass term of the Lagrangian as
\begin {equation}
   {\textstyle{1\over2}} \, m^2 \, \phi^2
   =
   {\textstyle{1\over2}} \, \meff^2 \, \phi^2
   + {\textstyle{1\over2}} \, (m^2{-}\meff^2) \, \phi^2 \,,
\end {equation}
treating the first term on the right-hand side as part of the
unperturbed Lagrangian and 
the second term as a perturbation.%
\footnote{
   See, for example, ref.~\cite{arnold&espinosa}.
}
This perturbation will generate a new graph at NLO, shown in
fig.~\ref{fig:counterterm}.

\begin {figure}
\vbox
    {%
    \begin {center}
	\figrule
	\epsfbox [180 340 480 460]{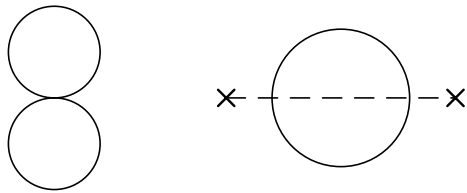}%
	\figrule
    \end {center}
    \caption
	{%
	\label {fig:2loop heavy}
        Two-loop diagrams contributing to $V_2^\heavy$.
	}%
    }%
\end {figure}

For the two-loop potential, first consider the $\eps \to 0$ and
$m \to 0$ limit of (\ref{eq:V2 form}), which we now refer to as
$V_2^\heavy$.  For the current calculation, we
will need to know all of the coefficients $\{ \, \vconst_{2i} \, \}$.
$\vconst_{21}$ may be determined by either explicit calculation or by applying
the two-loop RG to the one-loop potential (\ref{eq:1loop w u+v}).
The constant $\vconst_{20}$, however, requires explicit calculation.
The two-loop contributions are given the diagrams of fig.~\ref{fig:2loop heavy}
combined with fig.~\ref{fig:counterterm}.
\noindent
The details of the calculation
are given in Appendix~\ref{apndx:two-loop}, and one finds
   \begin {equation}
      \vconst_{21} = -{(n{+}6)\over3} \, \vconsta \,,
      \qquad
      \vconst_{20} =  {(n{+}18)\over6} \, \vconsta \,,
   \label{eq:vconst2j}
   \end {equation}
in addition to the previous result (\ref{eq:vconst22}) for $\vconst_{22}$.

The sub-leading corrections to $V_2^\heavy$ come from relaxing the
$\eps{\to}0$ limit
and remembering that the heavy mass (\ref{eq:mb def})
is $M^2+m^2$ instead of simply $M^2$.
Expanding in $\eps$ and $m^2/M^2$
gives
   \begin {eqnarray}
      \norm \mu^\eps \, \delta V_2^\heavy &=&
      u \, m^2 M^2 \left[
          {(n{-}1)\over3}\, \vconsta \ln^2 \!\left(M^2\over\mu^2\right)
          - {(n{+}4)\over3}\, \vconsta \ln \!\left(M^2\over\mu^2\right)
          + \vconst_{m20} \right]
   \nonumber\\ && {}
      + \eps \, u M^4 \Biggl[
          - {(n{+}2)\over12}\, \vconsta \ln^3 \!\left(M^2\over\mu^2\right)
          + {(3n{+}14)\over12}\, \vconsta \ln^2 \!\left(M^2\over\mu^2\right)
   \nonumber\\ && \qquad\qquad\qquad\qquad\quad
          - \left({n{+}10\over3} + {(n{+}2)\over36}\pi^2\right)
                    \vconsta \ln \!\left(M^2\over\mu^2\right)
          + \vconst_{\eps 20} \Biggr] \,,
   \label{eq:del V2}
   \end {eqnarray}
where the coefficients of all the logs can be determined by requiring the
full potential to be invariant under the two-loop RG.
The constant $\vconst_{\eps 20}$ will be
irrelevant as it can be absorbed into a redefinition of $\mu$.
The remaining constant, $\vconst_{m 20}$, is found by expansion of the
explicit ($\eps{=}0$) two-loop potential, as described in
Appendix~\ref{apndx:two-loop}.  One finds
\begin {equation}
   \vconst_{m 20} = -{1\over3} \, \vconsta \,.
\end {equation}

Finally, the contribution of heavy modes to the three-loop
potential has the form
\begin{equation}
   \normu V_3 = u^2 M^4 \left[
        \vconst_{33}\ln^3\!\left(M^2\over\mu^2\right)
      + \vconst_{32}\ln^2\!\left(M^2\over\mu^2\right)
      + \vconst_{31}\ln\!\left(M^2\over\mu^2\right)
      + \vconst_{30} \right]
      + O(\eps^3 V)_\asym \,.
\label{eq:V3 form}
\end{equation}
We will only need $\vconst_{33}$ and $\vconst_{32}$ for
$\chi_+/\chi_-$.
($\vconst_{31}$ can be changed by a suitable redefinition of the couplings,
and so cannot affect the physical ratio $\chi_+/\chi_-$.)
The coefficients $\vconst_{33}$ and $\vconst_{32}$
can be determined by applying the two-loop RG to the full potential.
To do so, we first need to relax the restriction $u{=}{-}v$
in our calculation of the two-loop potential.  The analysis can
be simplified a bit, however, in that we can treat $u{+}v$ as small and
only keep terms linear in $u{+}v$.  (That is because the RG derivative
$\beta_u \partial_u + \beta_v \partial_v$ acting on $u{+}v$ does not give
zero when $u = -v$, although the same operation on $(u{+}v)^2$ does yield zero.)
Keeping only such terms in the $m{=}0$, $\eps{\to}0$ approximation to the
potential gives the following analog to (\ref{eq:1loop w u+v}):
\begin {eqnarray}
   \norm \mu^\eps (V_0 {+} V_1 {+} V_2) &=&
   \Lambda +
   M^4 \Biggl\{
       \left[ \vconsta \ln\!\left(M^2\over\mu^2\right)\!
	       + \vconstb
       \right]\!
   + u \left[
	   \vconst_{22} \ln^2\!\left(M^2\over\mu^2\right)\!
	   + \vconst_{21} \ln\!\left(M^2\over\mu^2\right)\!
	   + \vconst_{20}
     \right]
\nonumber\\ && \qquad\qquad {}
   + {3\over2} {(u{+}v) \over u^2}\!
   + (u{+}v) \!\left[
          -{\vconsta \over 2} \ln^2\!\left(M^2\over\mu^2\right)\!
          - 3\,\vconsta \ln\!\left(M^2\over\mu^2\right)\!
          - {5\,\vconsta \over 2}
       \right]
\nonumber\\ && \qquad\qquad {}
   + O\bigl((u{+}v)^2\bigr) + O(\eps) 
   \Biggr\} .
\label{eq:u+v expansion}
\end {eqnarray}
Combining with $V_3$, applying the RG, and then taking $u{=}{-}v$ yields
\begin {equation}
   \vconst_{33} = {\textstyle {1\over36}} \,
		   (n^2 + 8) \, \vconsta \,,
   \qquad
   \vconst_{32} = -{\textstyle {1\over36}} \,
		   (2 n^2 +45 n +7 ) \, \vconsta \,.
\end {equation}


\subsection{The effective potential: symmetric phase}

In order to find the critical value of $m^2$, we must equate the
free energy of the two phases.  The asymmetric phase approximation
(\ref{eq:V asym}) is not good in the $M{=}0$ symmetric phase because it
relies on the approximation $m^2/M^2 \sim O(\eps)$.  We must compute
the symmetric-phase free energy independently.  At the order of interest,
it is just the one-loop contribution (\ref{V1 with m}):
\begin {equation}
   \normu V(0) = \Lambda
        + {n\over4} \, m^4 \left[
              \ln\!\left(m^2\over\mu^2\right) - {3\over2} \right]
        + O\bigl(\eps^3 M_1^4\bigr) \,.
\end {equation}

For the NNLO susceptibility in the symmetric phase, we need the curvature
of the two-loop potential at $M{=}0$.  This is obtained by differentiating
the general result (\ref{full V2}) of Appendix~\ref{apndx:two-loop},%
\footnote{
  Replace $J$ and $I$ by $\hat J$ and $\hat I$ in (\ref{full V2})
  and remove $1/\eps$ poles, as discussed in the appendix.
}
and yields (\ref{eq:NNLO sym curv}).


\subsection{$(u_*,v_*)$ at NLO}
\label{sec:ustar2}

The NNLO result for $\chi_+/\chi_-$ will require a NLO value for
$(u_*,v_*)$.
Consider the one-loop result for the RG trajectories,
(\ref{trajectories}) with $c=0$ or $1$,
flowing from the tricritical points.  Consider the solution for $u$ as a
function of $f\equiv u/v$, and call it $u^{[1]}(f)$.
Now look at the perturbation as we include higher loops.
Start by making the rescaling
\begin {equation}
   (u,v) = (\eps \, \bar u, \eps \, \bar v) \,,
   \qquad
   \mu = \bar\mu^{-1/\eps} \,,
\label{eq:scale dim}
\end {equation}
which makes the
$\eps$ expansion explicit in the RG equations:
\begin {mathletters}
\begin {eqnarray}
   \bar\mu {\partial \bar u \over \partial {\bar\mu}} &=&
     - \bar u + \beta_u^{(1)}(\bar u,\bar v)
     + \eps\, \beta_u^{(2)}(\bar u, \bar v) + \cdots \,,
\\
   \bar\mu {\partial \bar v \over \partial {\bar\mu}} &=&
     - \bar v + \beta_v^{(1)}(\bar u,\bar v)
     + \eps\, \beta_v^{(2)}(\bar u, \bar v) + \cdots \,,
\end {eqnarray}
\end {mathletters}
where $\beta^{(n)}$ is the $n$-th order contribution to the $\beta$-function.
Now expand
\begin {equation}
   \bar u(f) = \bar u^{[1]}(f) + \eps\,\delta(f) + O(\eps^2) \,.
\end {equation}
Plugging into the renormalization group equations, linearizing in the
perturbation $\delta$, and solving yields
\begin {equation}
   \delta(f) = e^{K(f)} \int\nolimits_{f_0}^f df' e^{-K(f')}
      \left.\left[ {\beta_u^{(2)} \over \beta_f^{(1)}}
           - {\beta_f^{(2)} \beta_u^{[1]} \over (\beta_f^{(1)})^2} \right]
      \right|_{\bar u^{[1]}(f'), f'}
       + e^{K(f)} \delta(f_0) \,,
\label{eq:deltaf}
\end {equation}
where
\begin {eqnarray}
   K(f) &\equiv& \int\nolimits_{f_0}^f df' \left.
     {\partial \over \partial u} \left(\beta_u^{[1]}\over\beta_f^{(1)}\right)
   \right|_{\bar u^{[1]}(f'), f'} \,,
\\
   \beta_f &\equiv& {1\over v} \, \beta_u - {u\over v^2} \, \beta_v \,,
\label{eq:betaf}
\\
\noalign {\hbox {and we have defined}}
   \beta_u^{[1]} &\equiv& -u + \beta_u^{(1)} \,.
\end {eqnarray}
For the trajectory flowing from the Ising fixed point, we should find
that the dependence on the initial perturbation $\delta(f_0)$ vanishes
when $f_0$ approaches the Ising tricritical line $f_0{=}0$.  In this case, the
system will first flow to the Ising fixed point before flowing away, thus
washing away dependence on $\delta(f_0)$.
Similarly, for flow from the cubic fixed point,
the dependence on $\delta(f_0)$ should vanish as $f_0 \to -\lambda$.
This is indeed the case.  For any trajectory,
one finds
\begin {equation}
   e^{K(f)} = \left(f_0\over f\right)^2
          \left(f_0+\lambda\over f+\lambda\right)^{n\lambda} \,,
\end {equation}
which vanishes as $f_0 \to 0$ or $-\lambda$.

The resulting correction
$\delta(f)$ does not seem to have a simple form for general $n$,
but may be evaluated explicitly when $n=2$.
Taking $f_0=-\lambda$ and $f=-1$ for the
cubic point trajectory, one finds
\begin {equation}
   (u_*,v_*) = (u_*,-u_*) \,,
   \qquad
   u_* = \textstyle{51\over80} \, \eps
     + \left( \textstyle{243\over80} \ln\textstyle{3\over2}
             - \textstyle{171\over200} \right) \eps^2
     + O(\eps^3) \,.
\label{eq:NLO ustar}
\end {equation}
One may verify that the analogous calculation for the Ising fixed point
trajectory yields the appropriate transformation (\ref{uv map})
of (\ref{eq:NLO ustar}).


\subsection{Results}

By equating the potential in the symmetric and asymmetric phases, one
determines $m^2$ and the asymmetric phase value of $M^2$ to
next-to-next-to-leading order, on the line $u{+}v = 0$.
One finds
\begin {eqnarray}
      m^2 &=& m_1^2 \, \Biggl\{ \, 1
        + \eps \left[{9\over16} + {\pi^2\over12}\right]
        + u \left[{n\over24} - {13\over6}\right]
   \nonumber\\ && \;\qquad {}
        + \eps^2\left[-{269\over1536} - {\pi^2\over64} + {\pi^4\over288}
                - {\vconst_{\eps\eps10}\over\vconsta}\right]
        + \eps \, u \left[{27n^{\vphantom 2}\over128} + {41\over32} +
                \left({n\over32}-{1\over72}\right)\pi^2
                - {\vconst_{\eps20}\over\vconsta}\right]
   \nonumber\\ && \;\qquad{}
        + u^2 \left[{13 n^2\over1152}
		+ {19 n \over 72} + {1 \over 9}
                - {n{-}1\over36}\ln 2
                + {(n{-}1)(n{-}4)\over144} \,
		    \ln\!\bigg({n{-}1\over12}u\bigg)\!
                - {\vconst_{31}\over2\vconsta}
                - {\vconst_{30}\over\vconsta}
                \right]
   \nonumber\\ && \;\qquad {}
      + O(\eps^3) + O(u^3) \Biggr\} \,,
   \end {eqnarray}
and
   \begin {eqnarray}
      M^2 &=& M_1^2 \, \Biggl\{ \, 1
        + \eps\left[{1\over16} + {\pi^2\over12}\right]
        + u\left[-{n\over24} - {1\over4}\right]
   \nonumber\\ && \!\!\!\qquad {}
        + \eps^2\left[-{701\over1536} - {11\pi^2\over192} + {\pi^4\over288}
                - {\vconst_{\eps\eps10}\over\vconsta}\right]
        + \eps \, u\left[{47n\over384} + {247\over64} 
                + {7(n{+}6)\over288}\pi^2
                - {\vconst_{\eps20}\over\vconsta}\right]
   \nonumber\\ && \!\!\!\qquad {}
        +  u^2 \!\left[{49n^2\over1152}
		+ {203 n \over 96} - {101 \over 144}
                - {n{-}1\over6}\ln 2
                - {(n{-}1)(n{+}24)\over144}
		    \ln\!\left( {n{-}1\over12}u\right)\!
                - {3\over2} {\vconst_{31}\over\vconsta}
                - {\vconst_{30}\over\vconsta}
                \right]\!
   \nonumber\\ && \!\!\!\qquad {}
      + O(\eps^3) + O(u^3) \Biggr\} \,.
\end {eqnarray}
The curvatures of the potential in the two phases, at the transition,
when $u{+}v=0$, are
   \begin {eqnarray}
      \partial_\phi^2 V(0) &=& m^2 \, \Biggl\{ 1
        + u \, {(n{-}1)\over6} \! \left[
             \ln\!\left({n{-}1 \over 12} u\right)
             - {\textstyle{1\over2}}  \right]
        +  \eps \, u \, {(n{-}1)\over24} \! \left[
             - \ln^2\!\left({n{-}1\over 12} u\right)
             + \ln\!\left({n{-}1\over 12}u\right)
             + 1  \right]
   \nonumber\\ && \!\qquad {}
        + u^2 \, {(n{-}1)\over36} \!\left[
            (n{+}2) \ln^2\!\left({n{-}1\over 12}u\right)\!
            - 9 \ln\!\left({n{-}1\over 12}u\right)\!
            - 12\sqrt3 \left({\pi\over6}\ln2
		- L\!\left(\pi\over6\right)\!
		\right)\!
            - 3
        \right]\!
   \nonumber\\ && \qquad {}
      + O(\eps^3) + O(u^3) \Biggr\} \,,
   \label{eq:NNLO sym curv}
   \\
      \partial_\phi^2 V(\phi) &=& 2 m^2 \, \Biggl\{ 1
        - \half \, \eps
        - u\, {(n{-}13)\over12}
        - \eps \, u \, {(n{-}9)\over12}
   \nonumber\\ && \;\qquad {}
        + u^2 \left[
            {(n{-}1)(27-2n)\over72}\,
	    \ln\!\left({n{-}1\over12}u\right)
            + {3(n{-}1)\over8} \ln2
            - {(103n-73)\over72}
        \right]
   \nonumber\\ && \;\qquad {}
      + O(\eps^3) + O(u^3) \Biggr\} \,.
   \end {eqnarray}
Here, $L(z)$ is Lobachevskiy's function, defined in
Appendix~\ref{apndx:two-loop}, and
\begin {equation}
   L\!\left(\pi\over6\right) = 0.02461715\cdots \,.
\label{eq:lob value}
\end {equation}
As required, all dependence
on the undetermined parameters $\vconst_{30}$, $\vconst_{31}$, {\em etc}.,
is hidden in the overall factor of $m^2$.
The susceptibilities $\chi_+$ and $\chi_-$ equal these curvatures
up to a common overall proportionality constant.
Inserting the value of $u_*$ derived in the previous section
yields our
final result for the susceptibility ratio at $n{=}2$,
\begin {eqnarray}
   {\chi_+\over\chi_-} &=& 2 \, \Biggl\{
     1
     + \eps \left[
	 - {17\over160}\, \ln\!\left(17\eps\over320\right)
	 + {11\over80}
	 \right]
\nonumber \\ && \qquad {}
     + \eps^2 \left[
	 - {187\over160^2}\, \ln^2\!\left(17\eps\over320\right)
	 + \left( {8374\over160^2}
	        - {81\over160} \, \ln\!\left(3\over2\right)\right)
	     \ln\!\left(17\eps\over320\right)
	\right.
\nonumber \\ && \;\qquad\qquad {}
     - \left.
         {55129\over 2\cdot 160^2}
         + {867 \sqrt3 \over 40\cdot 160} \left(
                {\pi\over6}\ln2 - L\!\left(\pi\over6\right) \right)
         + {81\over32} \, \ln\!\left(3\over2\right)
         + {7803 \, \ln 2 \over2\cdot160^2}
     \right]
\nonumber \\ && \qquad {}
     + O(\eps^3)
   \Biggr\} \,,
\label{eq:chi ratio result}
\end {eqnarray}


\section {Discussion}
\label{sec:discussion}

We now collect our results for $n{=}2$ and evaluate the coefficients
numerically:
\begin {eqnarray}
   {\chi_+\over\chi_-} &=&
   2 \, \Bigl[ 1 + \eps \, (-0.1063\ln\eps + 0.4494) 
\\ && \quad\; {}
      + \eps^2(-0.0073\ln^2\eps + 0.1647\ln\eps - 0.2859)
      + O(\eps^3) \Bigr] ,
\nonumber
\\
   {\xi_+^2\over\xi_-^2} &=&
   2 \, \Bigl[ 1 + \eps \, (-0.1063\ln\eps + 0.4139) 
      + O(\eps^2) \Bigr] .
\end {eqnarray}
The ratio $C_+/C_-$ of specific heats will be evaluated in
ref.~\cite{specific-heat}, and
all of these results are compared against Monte Carlo simulations
\cite{numerical} in ref.~\cite{summary}.

Looking solely at the series above, our results are moderately encouraging.
At $\epsilon {=} 1$,
the NLO corrections are 45\% for $\chi_+/\chi_-$ and 41\%
for $\xi^2_+ / \xi^2_-$.
NNLO corrections for $\chi_+/\chi_-$ drop to 29\%.
The subleading corrections could have turned out quite large compared to
the leading-order results, as is known to happen in some cases where
the leading-order $\epsilon$ expansion is a poor quantitative
approximation.%
\footnote{
   See, for example, the discussion of large $n$ in ref.~\cite{arnold&yaffe1}.
}
In contrast, our series seem tolerably well behaved.

It should not be too difficult a calculation to extend $\xi_+^2/\xi_-^2$
to NNLO, but we have not done so.
An interesting question for further research is whether
it is possible to determine the large-order behavior of $\eps$ expansions
for the ratios we have investigated.
The techniques used for critical exponents of
second-order transitions%
\footnote{
  For a review, see secs.~27.3 and 40 of ref.~\cite{zinn-justin} and
  references therein.
}
\cite{large-order}
do not obviously generalize to this problem.


\bigskip

This work was supported by the U.S. Department of Energy
grants DE-FG06-91ER40614 and DE-FG03-96ER40956.
We thank David Broadhurst and Joseph Rudnick for useful conversations.


\appendix

\section {Review of leading-order \protect\lowercase{$(u_*,v_*)$}}
\label{apndx:ustar}

Consider the one-loop RG equation given by (\ref{eq:uv RG1}) and
(\ref{eq:beta1}).  Before we look for trajectories, note
the location of the fixed points at leading order in $\eps$:
\begin {eqnarray}
  \hbox{Gaussian:} &&\qquad  (u,v) = (0,0) \,,                             \\
  \hbox{Ising:   } &&\qquad  (u,v) = (0, \textstyle{\eps\over3}) \,,      \\
  \hbox{\On:  }    &&\qquad  (u,v) = (\textstyle {3 \eps\over(n{+}8)}, 0) \,,\\
  \hbox{cubic:   } &&\qquad  (u,v) = (\textstyle {\eps\over n},
				      {(n{-}4)\over3n}\eps) \,.
\end {eqnarray}
Though we shall not directly make use of it now, note that the dependence
on dimension in the one-loop RG equations can be eliminated by the rescaling
(\ref{eq:scale dim}) so that (\ref{eq:uv RG1}) becomes
\begin {mathletters}%
\label{eq:uv-bar RG1}%
\begin {eqnarray}
   \bar\mu{\partial \bar u \over \partial {\bar\mu}}
   &=& -\bar u + \betaone_u(\bar u, \bar v) \,,
\\
   \bar\mu {\partial \bar v \over \partial {\bar\mu}}
   &=& -\bar v + \betaone_v(\bar u, \bar v) \,.
\end {eqnarray}%
\end {mathletters}%
This depends on the fact that the $\betaone$ are quadratic in $u$ and $v$.

The following is a sketch of the solution to the RG equations
(\ref{eq:uv RG1}) as found
by Rudnick \cite{rudnick} and Domany, {\it et al.} \cite{domany}.
Begin by removing the first term on the right-hand side
by switching to new variables $A$ and $B$,
\begin {equation}
   u = \mu^{-\eps} A \,,
   \qquad
   v = \mu^{-\eps} B \,,
\end {equation}
so that
\begin {eqnarray}
   \mu^{1+\eps} {\partial A \over \partial \mu} &=& \betaone_u(A, B) \,,
\qquad
   \mu^{1+\eps} {\partial B \over \partial \mu} = \betaone_v(A, B) \,.
\end {eqnarray}%
Divide these two equations, and note that the right-hand side is a
function solely of $f \equiv u/v = A/B$:
\begin {equation}
   {d A\over d B} =
   {\betaone_u(A, B) \over \betaone_v(A,B)}
   \equiv H(f) \,.
\end {equation}
Changing variables from $(A,B)$ to $(f,B)$, solving the resulting equation,
and fixing the boundary condition $B(f_0) = B_0$ gives
\begin {equation}
   B = B_0 \exp \left(\int\nolimits_{f_0}^f {df' \over H(f')-f'} \right) \,.
\end {equation}
With the beta functions (\ref{eq:beta1}), we have
\begin {mathletters}
\label{eq:uv solns}
\begin {eqnarray}
   {\mu^\eps \, u\over\mu_0^\eps \, u_0} &=&
                     \left(f_0\over f\right)^2
                     \left(f_0 + \lambda \over f+\lambda\right)^{n\lambda} \,,
\\
   {\mu^\eps \, v\over\mu_0^\eps \, v_0} &=&
                     \left(f_0\over f\right)^3
                     \left(f_0 + \lambda \over f+\lambda\right)^{n\lambda} \,,
\label{eq:uv soln b}
\end {eqnarray}
\end {mathletters}
with $\lambda$ given by (\ref{eq:lambda}).
Now, from $f=u/v$, note that
\begin {equation}
  \mu {\partial f \over \partial \mu} =
  {1\over v} \, \betaone_u - {u\over v^2} \, \betaone_v \,.
\end {equation}
Use the solutions (\ref{eq:uv solns}) to write the right-hand side in
terms of $f$, $\mu$, and constants.  Solving the resulting differential
equation yields
\begin {equation}
   \mu_0^\eps \left[1 - {\eps\over v} R(f,c)\right] =
     \mu^\eps \left[1 - {\eps \over v_0} R(f_0,c) \right] \,,
\label{eq:mu soln}
\end {equation}
where
\begin {equation}
   R(f,c) \equiv
      {\lambda^2 \over (n\lambda+1)(n\lambda+2)} {1\over f^3}
      \left[ {(n\lambda+1)\over\lambda} f^2 - 2 f
      + {2\over n}
      - {2c\over n} \left(1 + {f\over\lambda}\right)^{-n\lambda}
      \right] .
\end {equation}
The equation (\ref{eq:mu soln}) is independent of the constant
$c$ by (\ref{eq:uv soln b}).  We have introduced $c$ as a trick for finding
the final equation for all the trajectories.  One way (\ref{eq:mu soln})
can be solved is to have
\begin {equation}
   1 - {\eps\over v} R(f,c) = 0
\end {equation}
for all $f$ and $v$ on the trajectory.  Different $c$ correspond to
different solutions and give all possible trajectories (either directly or
as limiting cases).
%


\section {Details of two-loop potential}
\label{apndx:two-loop}

\begin {figure}
\vbox
    {%
    \begin {center}
	
	\figrule
	\epsfbox [100 335 510 445] {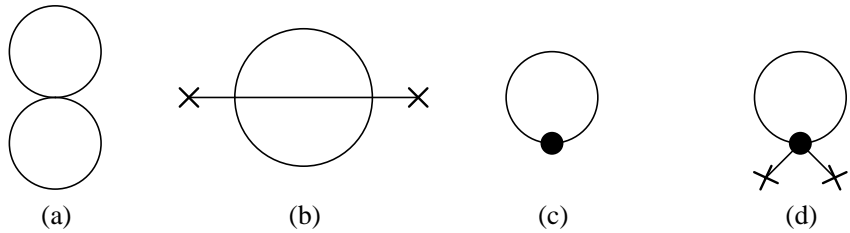}
	\figrule
    \end {center}
    \caption
	{%
	\label {fig:2loop}
        Two-loop diagrams contributing to the effective potential.
	The heavy dots represent renormalization counterterms.
        Each line represents both heavy and light mode contributions.
	}%
    }%
\end {figure}

   The two-loop diagrams (a) and (b) of fig.~\ref{fig:2loop}
give the following contribution to
the two-loop potential:
\begin {mathletters}%
\label {full V2}%
\begin {eqnarray}%
   V_2^{\rm (a)} &=& {1\over8} \norm (u+v) J(\ma^2,\ma^2)
      + {1\over12} (n{-}1) \norm u J(\ma^2,\mb^2)
\nonumber\\ && \qquad
      + {1\over8}(n{-}1) \norm (u+v) J(\mb^2,\mb^2)
      + {1\over24} (n{-}1)(n{-}2) \norm u J(\mb^2,\mb^2) \,,
\\
   V_2^{\rm (b)} &=& - \norm^2\phi^2 \left[
      {1\over12} (u+v)^2 I(\ma^2,\ma^2,\ma^2)
      + {(n{-}1)\over36} u^2 I(\mb^2,\mb^2,\ma^2)
   \right] \,,
\end {eqnarray}%
\end {mathletters}%
where
\begin {eqnarray}
   J(x,y) &=& J(x) J(y) \,,
\\
   J(x) &=& {1\over(2\pi)^d} \int {d^dk \over k^2+x}
   = {1\over(4\pi)^{d/2}} \Gamma\left(1-{d\over2}\right)
       x^{{d\over2}-1} \,,
\\
\noalign {\hbox{and}}
   I(x,y,z) &=& {1\over(2\pi)^{2d}}
     \int {d^dk \, d^dq \over (k^2+x)\, (q^2+y)\, [(k+q)^2+z]} \,.
\end {eqnarray}
Contributions were considered from all mixtures of
light ($m_a$) and heavy ($m_b$) lines, with masses given by
(\ref{eq:mab defs}).
The effect of diagrams (c) and (d) involving one-loop counter-terms
is to replace
$J(x,y)$ and $I(x,y,z)$ in (\ref{full V2}) by
\begin {eqnarray}
   \norm J (x,y) \to
   \norm \hat J(x,y)
     &\equiv& \norm J(x,y) + {2\over\eps} \, [x J(x) + y J(y)] \,,
\\
   \norm^2 I(x,y,z) \to
   \norm^2 \hat I(x,y,z)
     &\equiv& \norm^2 I(x,y,z) - {2\over\eps} \, \norm [J(x)+J(y)+J(z)] \,.
\end {eqnarray}
Including two-loop counter-terms corresponds to simply throwing away
any remaining $1/\eps$ and $1/\eps^2$ pieces in the potential.%
\footnote{
   Note that the relationship between $M$ and $\phi$ has non-trivial
   dependence on $\eps$ through $\norm$.
   Including the two-loop counterterms corresponds to throwing away
   remaining terms of the form
   $\norm^{a-1}\eps^{-n} u^3 m^{2(2-a)} \phi^{2a}$ in $V$ or
   $\eps^{-n} u^{3-a} m^{2(2-a)} M^{2a}$ in $\normu V$.
}

  The general form for $I(x,y,z)$ for arbitrary arguments in arbitrary
dimension is given in ref.~\cite{ford}.  We shall need only the following
special cases for the various expansions we make:%
\footnote{
    Note that our $\eps$ is $4{-}d$ where as that of ref.~\cite{ford} is
    $(4-d)/2$.
    Also \cite {ford-private}, there are some typographical errors in
    ref.~\cite{ford}.
    In their equations (5.8--15), each explicit factor
    of $J(w)$ in those equations (but not $J(v,w)$ or $\hat J(v,w)$)
    should be multiplied by $1/\kappa$.  The factors of $\mu^{2\bar\eps}$
    in (3.4--5) should be eliminated.  The left-hand side of (4.8)
    should be $x^2{+}y^2{+}z^2$.  The second term on
    the right-hand side of (4.26) should be multiplied by 2.
}
   \begin {eqnarray}
      I(x,x,0) &=& {1\over(4\pi)^d} {\Gamma\!\left(2-{d\over2}\right)
        \Gamma\!\left(1-{d\over2}\right) \over d-3} x^{d-3} \,,
   \\
      \norm^2 I(x,x,x) &=& x \Biggl\{
        - {6\over\eps^2}
        + {1\over\eps}\left(6 \ln x - 9\right)
        - 3 \ln^2 x
        + 9 \ln x
   \nonumber\\ && \qquad
        - 6\sqrt3 \left[ L\left(\pi\over6\right) - {\pi\over6} \ln2 \right]
        - {21\over2} - {\pi^2\over2}
      \Biggr\}
      + O(\eps) \,,
   \end {eqnarray}
where $L(z)$ is Lobachevskiy's function, defined by
\begin {equation}
   L(z) \equiv - \int\nolimits_0^z dx \, \ln \, \cos x \,,
\end {equation}
and the value of interest is given in eq.~(\ref{eq:lob value}).

Ignoring the light mass $\ma$ in (\ref{full V2}) and expanding to
leading-order in $\eps$ with $u{=}{-}v$ gives:
\begin {eqnarray}
   V_2^\heavy &=&
   {(n{-}1)\over144} \, \norm u^2 \, \phi^2 \, \mb^2 \left[
      {(n{+}2)\over6} \ln^2\!\left(\mb^2\over\mu^2\right)
      - {(n{+}6)\over3}\ln\!\left(\mb^2\over\mu^2\right)
      + {(n{+}18)\over6}\right]
\nonumber\\ &&\qquad {}
      + O(\eps V_2)_\asym \,.
\label{eq:V2 heavy}
\end{eqnarray}
The coefficients $\vconst_{2j}$ of (\ref{eq:vconst22}) and (\ref{eq:vconst2j})
and $\vconst_{m20}$ of (\ref{eq:delta2 V1}) and (\ref{eq:vconstcm})
may then be extracted.
If one makes the above expansion without assuming $u{+}v$ is precisely zero
(and expands to first order in
$\ma^2 = {\textstyle {1\over2}} \normu (u+v) \, \phi^2 + O(m^2)$),
one obtains (\ref{eq:u+v expansion}).


\begin {references}

\bibitem {rudnick}
    J. Rudnick, {\sl Phys.\ Rev.\ B} {\bf 11}, 3397 (1975).

\bibitem {amit}
    D. Amit, {\it Field Theory, the Renormalization Group, and Critical
    Phenomena,} revised second edition (World Scientific, Singapore,
    1984);
    A. Aharony in {\sl Phase Transitions and Critical Phenomena: Vol.\ 6},
    eds. C. Domb and M. Green (Academic Press, 1976),~357.

\bibitem {summary}
    P. Arnold, S. Sharpe, L. Yaffe and Y. Zhang,
    University of Washington preprint UW/PT-96-25, in preparation.

\bibitem {alford}
    M. Alford and J. March-Russel, {\sl Nucl.\ Phys.}\ {\bf B417}, 527 (1994).

\bibitem {arnold&yaffe1}
    P.~Arnold and L.~Yaffe,
    {\sl Phys.\ Rev.\ D} {\bf 49}, 3003 (1994);
    University of Washington preprint UW/PT-96-28 (errata).

\bibitem {numerical}
    P. Arnold and Y. Zhang,
    University of Washington preprint UW/PT-96-26,
    hep-lat/96xxxxx.

\bibitem {specific-heat}
    P. Arnold and Y. Zhang,
    University of Washington preprint UW/PT-96-24,
    hep-ph/9610448.

\bibitem {3d}
    K. Farakos, K. Kajantie, M. Shaposhnikov,
      {\sl Nucl.\ Phys.} {\bf B425}, 67 (1994).

\bibitem {kettley}
    I. Ketley and D. Wallace,
    {\sl J. Phys.}\ {\bf A6}, 1667 (1973).

\bibitem {coleman&weinberg}
    S. Coleman and E. Weinberg,
    {\sl Phys.\ Rev.\ D} {\bf 7}, 1883 (1973).

\bibitem {domany}
    E. Domany, D. Mukamel, and M. Fisher,
    {\sl Phys.\ Rev.\ B} {\bf 15}, 5432 (1977).

\bibitem {arnold&espinosa}
    P. Arnold and O. Espinosa,
    {\sl Phys.\ Rev.\ D} {\bf 47}, 3546 (1993);
    {\bf 50} 6662(E) (1994).

\bibitem {zinn-justin}
    J. Zinn-Justin, {\sl Quantum Field Theory and Critical Phenomena},
    2nd edition (Clarendon Press, 1993).

\bibitem {large-order}
    E. Br\'ezin, J. Le Guillou, and J. Zinn-Justin,
    {\sl Phys.\ Rev.\ D} {\bf 15}, 1558 (1977).

\bibitem {ford}
    C. Ford, I. Jack and D. Jones, 
    {\sl Nucl. Phys.}\ {\bf B387}, 373 (1992).

\bibitem {ford-private}
    D. Jones, private communication.

\end {references}
\end {document}